\documentclass[journal]{IEEEtran}

\usepackage[english]{babel}
\usepackage{mathptmx}

\usepackage{amsmath}
\usepackage{graphicx}
\usepackage[colorlinks=true, allcolors=blue]{hyperref}
\usepackage{float}
\usepackage[normalem]{ulem}
\usepackage{cite}
\usepackage{amsfonts}

\begin{document}

\title{Reconfigurable Intelligent Surfaces as Spatial Filters} 

\author{{\IEEEauthorblockN{Sotiris Droulias, Giorgos Stratidakis, Emil Bj{\"o}rnson,~\IEEEmembership{Fellow,~IEEE} and \\ 
Angeliki Alexiou,~\IEEEmembership{Member,~IEEE}} }

\thanks{This work was supported by the European Commission’s Horizon Europe Programme under the Smart Networks and Services Joint Undertaking TERA6G project (Grant Agreement 101096949) and INSTINCT project (Grant Agreement 101139161). E.~Bj\"ornson was supported by the FFL18-0277 grant from SSF.}
\thanks{S.~Droulias, G.~Stratidakis and A.~Alexiou are with the Department of Digital Systems, University of Piraeus, 150, Androutsou St., 18534 Piraeus, Greece (Corresponding author: Sotiris Droulias, e-mail: sdroulias@unipi.gr).}
\thanks{E.~Bj\"ornson is with the Department of Computer Science, KTH Royal Institute of Technology, SE-16 440, Kista, Sweden.}
}

\maketitle

\begin{abstract}
The design of Reconfigurable Intelligent Surfaces (RISs) is typically based on treating the RIS as an infinitely large surface that steers incident plane waves toward the desired direction. In practical implementations, however, the RIS has finite size and the incident wave is a beam of finite $k$-content, rather than a plane wave of $\delta$-like $k$-content. To understand the implications of the finite extent of both the RIS and the incident beam, here we treat the RIS as a spatial filter, the transfer function of which is determined by both the prescribed RIS operation and the shape of the RIS boundary. Following this approach, we study how the RIS transforms the incident $k$-content and we demonstrate how, by engineering the RIS shape, size, and response, it is possible to shape beams with nontrivial $k$-content to suppress unwanted interference, while concentrating the reflected power to desired directions. We also demonstrate how our framework, when applied in the context of near-field communications, provides the necessary insights into how the wavefront of the beam is tailored to enable focusing, propagation with invariant profile, and bending, beyond conventional beamforming.
\end{abstract}


\section{Introduction}
\IEEEPARstart{A} Reconfigurable Intelligent Surface (RIS) reflects an incident wave ideally towards any desired angle \cite{Epstein2016, Tretyakov2020}. Owing to this unique feature, the RIS has recently attracted considerable attention as a means to create virtual line-of-sight (LOS) links to mediate non-LOS propagation in wireless communications. The need for increased data rates pushes communications systems to higher frequencies, particularly the millimeter-wave (mmWave) (30-100 GHz) and terahertz (THz) bands (0.1-10 THz) \cite{Withayachumnankul2018, Zhang2021}, where more spectrum is available. Since the coverage in these bands is mostly limited to LOS propagation, and the systems are vulnerable to blockage, the functionalities offered by the RIS are particularly useful for creating new strong LOS-like propagation paths.\\
\indent The principle of operation of the RIS is based on phase-shifting the reflected wave with respect to the incident wave, while preserving its magnitude. This is achieved by imposing a linear phase gradient on the incident wave; the RIS is composed of small, subwavelength elements (periodically distributed scatterers) that re-radiate the incident wave with prescribed phase advance, enabling reflection of the incident wave towards the desired direction. Based on this general principle, thus far several techniques have been theoretically proposed for the design of the desired RIS properties \cite{Capasso2011, Eleftheriades2014, Asadchy2016, Estakhri2016, Eleftheriades2016, Radi2017, Asadchy2017, Rubio2017, Eleftheriades2018, Rubio2019, Grbic2020, Rubio2021, DiRenzo2020, Alouini2021, Matthaiou2021} and relevant experiments have been performed in order to verify the predicted RIS performance \cite{Asadchy2017, Rubio2017, Eleftheriades2018, Rubio2019, Yang2016, Zhang2018, Dai2020, Bjornson2021, DeRosny2021}. The RIS is usually treated as a continuous surface, i.e., as a radiating sheet that locally satisfies the boundary conditions. This approach leads to the determination of continuous surface properties, such as surface impedance (or effective electric and magnetic surface conductivities), which ensure that an incident plane wave is reflected towards the desired direction upon impinging on the equivalent continuous surface; the surface properties are subsequently discretized, in essence rendering the continuous surface a collection of discrete, subwavelength scatterers \cite{Droulias2023}. \\
\indent For plane waves that illuminate surfaces of infinite extent, this general approach leads to an exact solution for the unknown properties of the RIS surface. However, in practical situations the RIS has finite size and, importantly, the incident waves are always beams of finite extent, rather than spatially completely delocalized plane waves. Therefore, the calculation of crucial quantities, such as the power at the position of a receiver, becomes a nontrivial task. The finite beam extent is associated with a finite but continuous interval of wavenumber content, referred to as the $k$-content,  as opposed to plane waves that correspond to a Dirac $\delta$-function in the $k$-space.
Taking this into account, in this paper we work in the $k$-space, in order to study how the RIS transforms the incident beam's $k$-content. To utilize this approach, we need to determine what is the RIS transfer function in the $k$-space and how it can be tailored to control the re-radiated wave. The transfer function establishes a unique correspondence between the incident and reflected $k$-components and, therefore, between the incident and reflected beam in real space. With this information, we can then answer crucial questions related to the impact of the RIS size, shape, and response on the incident beam, and to how good an approximation the plane wave approach is for practical situations. \\
\indent In this paper we work in the $k$-space to demonstrate how our understanding of the RIS operation can be significantly facilitated by examining the RIS as a spatial filter. Our novel approach offers an intuitive explanation to several RIS-related aspects, which are crucial for the quality of free-space LOS communication links. The main contributions are summarized as follows. 
\begin{itemize}
    \item The RIS operation is expressed analytically by a transfer function that transforms the incident beam's $k$-content.
    \item The reflected beam is expressed as a convolution of the incident beam's $k$-content and the RIS transfer function, taking into account the finite extent of both the beam's footprint and the RIS area.
    \item The far-field of the reflected beam is explained in terms of the balance between the contributions from the beam's $k$-content and the RIS transfer function. It is demonstrated that this balance is determined by the size of the incident beam's footprint relative to the RIS size.
    \item Beamsteering is treated as a shift operation in the $k$-space, explaining how it is possible that large reflection angles may lead to unintentional truncation of high $k$-components, thus affecting the power distribution of the reflected beam.     
    \item It is demonstrated how, by engineering the RIS transfer function, it is possible to manipulate the far-field distribution of the reflected wave to suppress undesired secondary lobes, while concentrating the power in desired directions. 
    \item It is demonstrated how the RIS response can be tailored to selectively suppress undesired incoming signals and transform a single incident beam into multiple beams towards multiple directions, with controllable gain towards selected angles.
    \item It is demonstrated how our framework can be utilized to optimize the RIS phase shifts, and to engineer beams that are capable of focusing, propagating with invariant profile and bending, beyond conventional beamforming.
\end{itemize}
\section{Spatial filtering}
\begin{figure}[t!]
\centering
    \includegraphics[width=\linewidth]{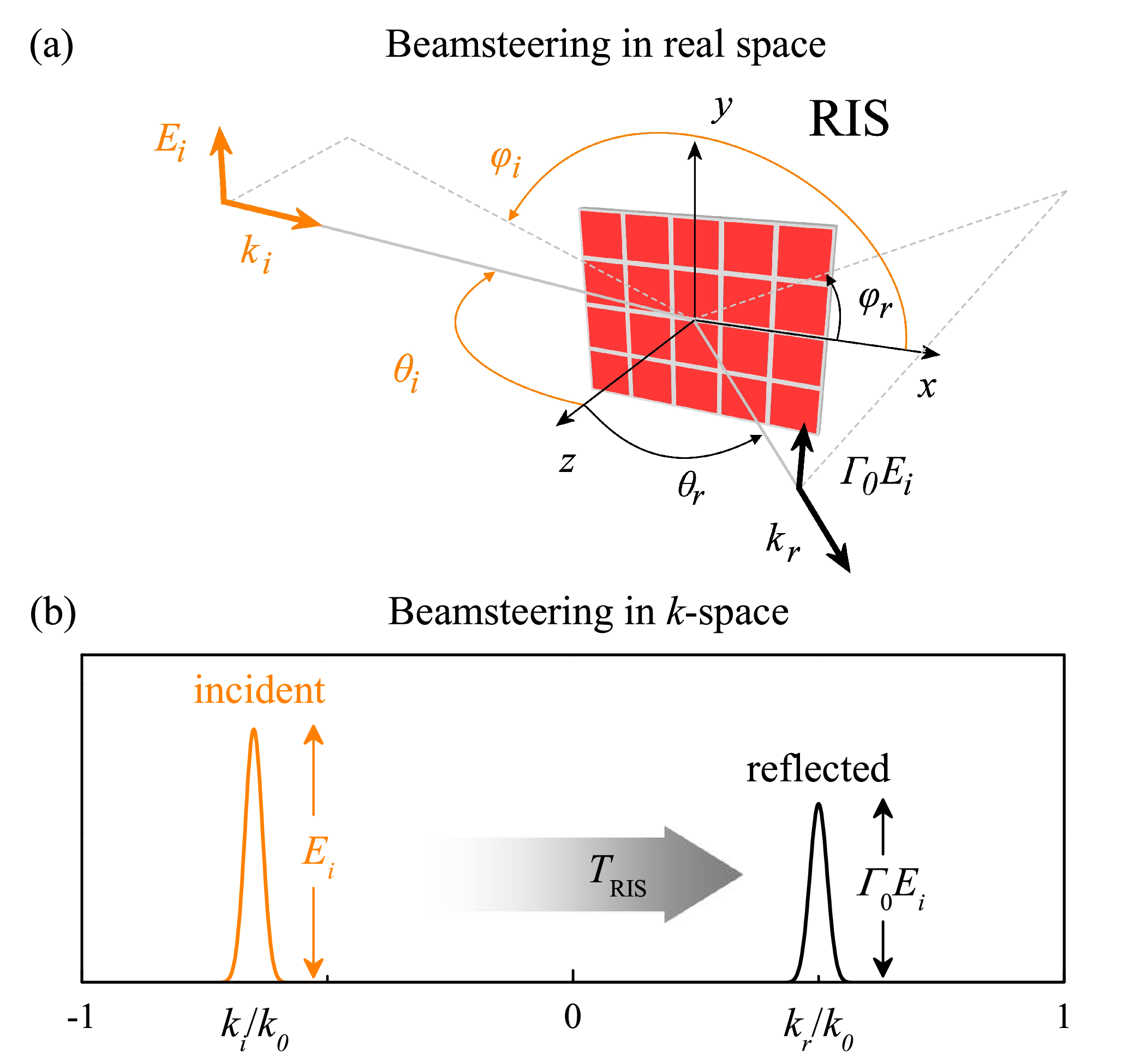}
	\caption{RIS-assisted beamsteering. (a) System setup, illustrating the angles of incidence ($\theta_i$,$\varphi_i$) and reflection ($\theta_r$,$\varphi_r$) at the RIS, associated with the wavevectors $\textbf{k}_i$ and $\textbf{k}_r$, respectively. (b) Beamsteering operation in the $k$-space. The RIS shifts the incident $k$-content by $k_r-k_i$ and scales its magnitude by $\Gamma_0$.}
    	\label{fig:fig01}
\end{figure}

\noindent Let us introduce our framework within the context of one of the most common RIS operations, namely beamsteering. As illustrated in Fig.\,\ref{fig:fig01}(a), the RIS is located at the origin of the global coordinate system and is illuminated by a TE-polarized wave, propagating in the $xz$-plane ($\varphi_i=\pi$, $\varphi_r=0$) at angle $\theta_i$, i.e., $\textbf{E}_i(\textbf{r}) = E_i(\textbf{r})\hat{\textbf{y}} \equiv  E_0 e^{-j \textbf{k}_i \textbf{r}} \hat{\textbf{y}}$, where $\textbf{k}_i = k_0 (\sin{\theta_i}\textbf{x}-\cos{\theta_i}\textbf{z})$ is the incident wavevector. The RIS steers the incident wave towards a desired angle $\theta_r$ and possibly modifies its magnitude, i.e., the reflected wave is written as $\textbf{E}_r(\textbf{r}) = E_r(\textbf{r})\hat{\textbf{y}} \equiv  \Gamma_0 E_0 e^{-j \textbf{k}_r \textbf{r}} \hat{\textbf{y}}$, where $\textbf{k}_r = k_0 (\sin{\theta_r}\textbf{x}+\cos{\theta_r}\textbf{z})$ is the reflected wavevector, $k_0=2\pi/\lambda$ the free-space wavenumber ($\lambda$ is the wavelength), and $\Gamma_0$ is a complex constant. The reflected $E$-field at the RIS plane ($z=0$) can also be written concisely as $E_r=\Gamma E_i$, where $\Gamma(x) = \Gamma_0e^{j(k_r-k_i)x}$ is the reflection coefficient, with $k_i=k_0\sin{\theta_i}$, $k_r=k_0\sin{\theta_r}$. The form of $\Gamma$ reveals that steering changes the in-plane $k$-component by an amount $k_0\sin{\theta_r}-k_0\sin{\theta_i}$; in the $k$-space this operation corresponds to simply shifting the in-plane $k$-content of the incident wave by $k_r-k_i$ (and scaling its magnitude by $\Gamma_0$), as illustrated in Fig.\,\ref{fig:fig01}(b). \\
\indent For beams of finite extent we can follow a similar path and express the reflected beam in the RIS plane as $E_r(x,y)=\Gamma(x,y)E_i(x,y)$, where $E_i(x,y)$ and $E_r(x,y)$ are the incident and reflected $E$-fields of the generalized beam in the RIS plane, respectively, and 
\begin{equation}
    \Gamma(x,y) = \Gamma_0(x,y)e^{j\phi(x,y)}
    \label{Eq:EqREFLCOEFF}
\end{equation}
expresses the local reflection coefficient. $\Gamma_0(x,y)$ is generally a real-valued function of space that accounts for changes in the magnitude of the incident wave, possibly including some global phase, which for simplicity we may ignore. The phase term $\phi(x,y)$ accounts for changes in the phase of the incident wave; for steering it takes the form $\phi(x,y) = (k_r-k_i)x=k_0(\sin{\theta_r}-\sin{\theta_i})x$. In the $k$-space, steering can then be equivalently described in terms of a transfer function, $T_\mathrm{RIS}$, that transforms the input $k$-spectrum of the incident beam according to
\begin{equation}
    \widetilde{E}_r (\textbf{k}) = \iint d\textbf{k}' \widetilde{T}_\mathrm{RIS}(\textbf{k},\textbf{k}')\widetilde{E}_i(\textbf{k}'),
    \label{Eq:EqErINT}
\end{equation}
where $\widetilde{E}_{i/r}(\textbf{k})=\widetilde{E}_{i/r}(k_x,k_y)$ is the Fourier transform of $E_{i/r}(x,y)$, and the tildes denote the functions in the Fourier space. It is important to note that the operation of the transfer function refers strictly to the wave transformation that occurs at $z=0$. However, because propagation in free space typically does not affect the $k$-content of the wave, the operation described by \eqref{Eq:EqErINT} provides the $k$-content of the reflected wave everywhere in $z>0$. The corresponding RIS transfer function is (see Appendix \ref{AppendixA} for derivation)

%
%

\begin{equation}
    \widetilde{T}_\mathrm{RIS}(k_x,k_y,k_x',k_y')=\widetilde{\Gamma}_0(k_x-k_x'+k_i-k_r,k_y-k_y'),
    \label{Eq:EqTRISa}    
\end{equation}
where $\widetilde{\Gamma}_0(k_x,k_y)$ is the Fourier transform of $\Gamma_0(x,y)$. The form of \eqref{Eq:EqTRISa} expresses the well-known modulation theorem in real-space \cite{Osgood2019}, and implies that the reflected wave results from convolving the incident $k$-content with the RIS impulse response, which is given by $\widetilde{\Gamma}_0$. Because $\Gamma_0(x,y)$ depends on the size and shape of the RIS (it can be considered practically zero outside the RIS area), the finite RIS extent has direct consequences on the reflected $k$-content. This becomes particularly apparent in the case of plane wave illumination, where $\widetilde{E}_i=\delta(k_x-k_i,k_y)$ and \eqref{Eq:EqErINT} yields
\begin{equation}
    \widetilde{E}_r(k_x,k_y) = \widetilde{\Gamma}_0(k_x-k_r,k_y).
\end{equation}
This implies that, for plane wave illumination, the $k$-content of the reflected field is just the Fourier transform of $\Gamma_0(x,y)$, shifted by $k_r-k_i$. This conclusion is also a good approximation for beams, as long as their $k$-contents are relatively narrow (see Appendix \ref{AppendixB} for details). Next, we consider beams with non-trivial $k$-content.

\subsection{Spatial filtering by an infinite RIS}

\begin{figure}[t!]
\centering
    \includegraphics[width=\linewidth]{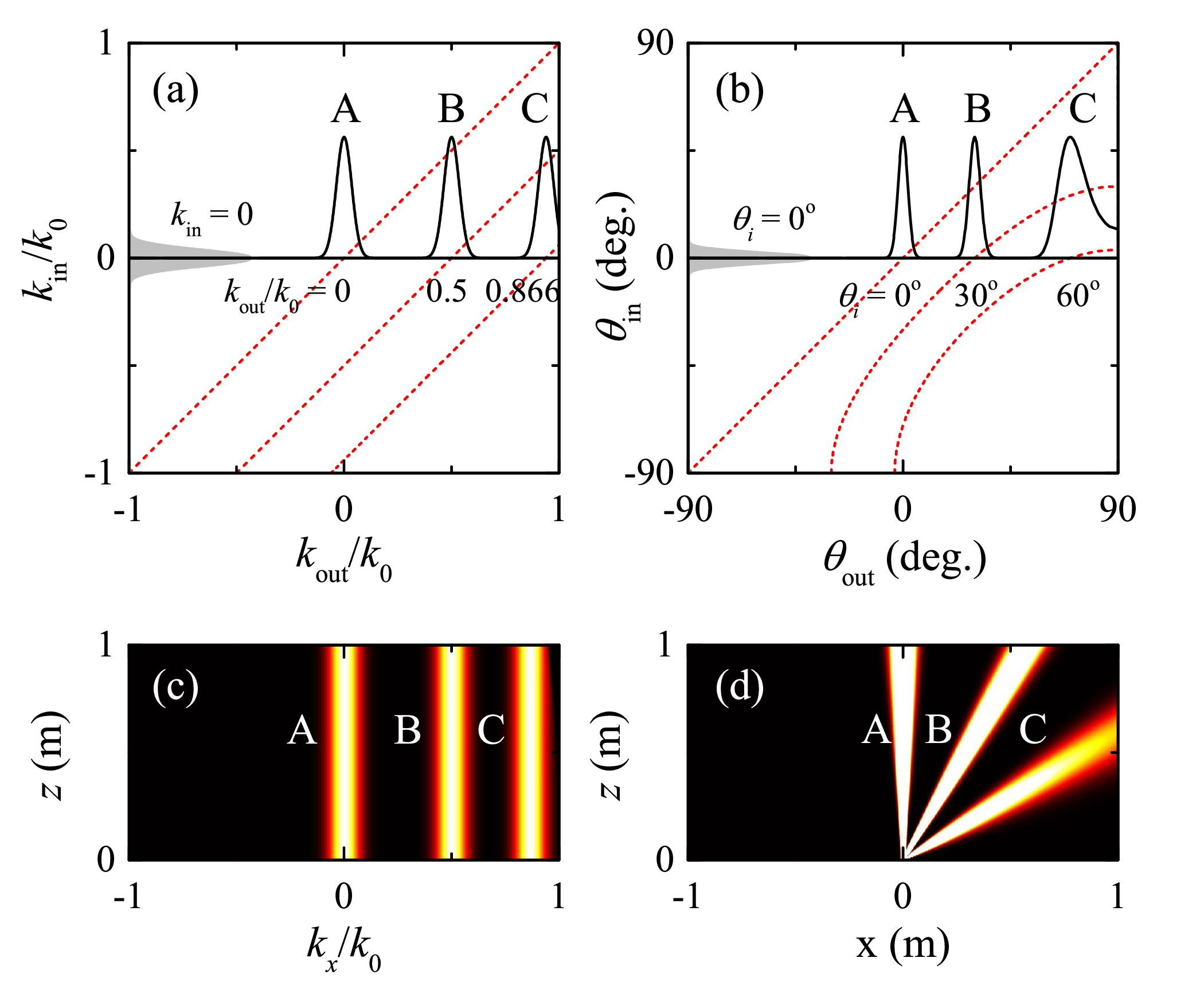} 
	\caption{The RIS transformation of the incident $k$-content expressed in terms of (a) wavenumbers and (b) angles. The red lines depict examples of the RIS transfer function that operates on the incident wave (solid gray) at normal incidence ($\theta_i=0^\circ$) for three steering angles, $\theta_r \in \{0^\circ,30^\circ,60^\circ\}$. The evolution of the reflected beams marked with the letters A,B and C is also shown in $k$-space (c) and in real space (d).}
    	\label{fig:fig02}
\end{figure}

\noindent For an infinitely large RIS, we may write $\Gamma_0(x,y) \equiv \Gamma_0$, where $\Gamma_0$ is a constant, as typically considered in many theoretical works \cite{Tretyakov2021}. In this case, the RIS transfer function takes the simple form
\begin{equation}
    \widetilde{T}_\mathrm{RIS}=\Gamma_0\delta(k_x-k_x'+k_i-k_r,k_y-k_y').
    \label{Eq:EqTRISb}
\end{equation}
Using \eqref{Eq:EqErINT}, we find the $k$-content of the reflected $E$-field as
\begin{equation}
    \widetilde{E}_r=\Gamma_0\widetilde{E}_i(k_x+k_i-k_r,k_y),
\end{equation}
i.e., the incident $k$-content is simply shifted by $k_r-k_i$ and scaled by $\Gamma_0$. This is illustrated in Fig.\,\ref{fig:fig01}(b), where the wavenumbers are normalized by $k_0$, to emphasize that the propagating components are restricted within the $|k_x/k_0|<1$ range (note that the wavenumber along the $z$-direction, $k_z=\sqrt{k_0^2-k_x^2}$, becomes imaginary, i.e., non-propagating, for $k_x>k_0$). \\
\indent As an example, in Fig.\,\ref{fig:fig02}(a) we consider a beam incident at $\theta_i$ and steered at $\theta_r \in \{ 0^\circ,30^\circ,60^\circ \}$. The red dashed lines denote $\widetilde{T}_\mathrm{RIS}$ as given by \eqref{Eq:EqTRISb} for each of the three steering operations, and the Gaussian distributions depict the $k$-content of the incident (filled gray) and reflected (black solid lines) beam. We see that, for large $\theta_r$, it is possible that a relatively wide $k$-content partially becomes non-propagating (see for example the case for $\theta_r=60^\circ$), contrary to the case of pure plane waves of $\delta$-like $k$-content, where the limit $|k_x/k_0|\rightarrow1$ can be approached arbitrarily close. The examples of Fig.\,\ref{fig:fig02}(a) can also be expressed in terms of the observation angle $\theta$ [Fig.\,\ref{fig:fig02}(b)], if we take into account that a component of the spatial spectrum of the field at spatial frequency $\sqrt{k_x^2+k_y^2}$ corresponds to a plane wave that is propagating at an angle of $\theta=\sin^{-1}(\sqrt{k_x^2+k_y^2}/k_0)$ to the $z$ axis. While both representations are equivalent, the representation in terms of $\theta$ provides insights into why a beam is expected to undergo increased spreading at larger steering angles $\theta_r$; this is a direct consequence of the coordinate stretching, when mapping the beam from the $k$-space to the $\theta$-space. Physically, what happens is that as the beam propagates, its $k$-content evolves with constant magnitude (see Fig.\,\ref{fig:fig02}(c)), however with phase that changes with propagation distance. In real space this manifests as beam diffraction, which becomes stronger at larger angles, as demonstrated in Fig.\,\ref{fig:fig02}(d), in accordance with well-known beam properties from phased array theory \cite{Aulock1960,Mailloux1982}. In these examples note how the peak of the beam drops with increasing beam width, as a consequence of the constant beam power.

%
%

\subsection{Impact of finite RIS size on spatial filtering}

\noindent Beams from real transmitters have finite extent and, as a result, the RIS surface is generally illuminated non-uniformly, i.e., the power density is not constant across the RIS. Depending on the incident beam's footprint, the RIS illumination area may vary from a small region up to the entire surface. For a RIS of finite size, the form of \eqref{Eq:EqTRISa} implies that the way that the $k$-content of the reflected beam scales depends strongly on the exact $k$-content of $\Gamma(x,y)$, which in turn depends on the spatial properties of the RIS. Therefore, the reflected beam can be tailored simply by engineering the RIS shape and size. \\
\indent For example, for a typical RIS of rectangular shape with the side lengths $L_x, L_y$, we may write $\Gamma_0(x,y) = \Gamma_0$ for $|x|<\frac{L_x}{2},|y|<\frac{L_y}{2}$, which leads to
\begin{equation}
    \widetilde{T}_\mathrm{RIS}=\Gamma_0 L_x L_y \textrm{sinc}(p_x)\textrm{sinc}(p_y),
    \label{Eq:EqTRISc}
\end{equation}
where $\textrm{sinc}(a)\equiv (\sin{a})/a$ and
\begin{subequations}
    \begin{gather}
        p_x = \frac{L_x}{2}(k_x-k_x'+k_i-k_r), \\
        p_y = \frac{L_y}{2}(k_y-k_y').
    \end{gather}
\end{subequations}
\noindent In essence, the $k$-content of the reflected field results from the convolution of the incident field's $k$-content with the Fourier transform of a rectangular aperture of size $L_x \times L_y$ and, therefore, it will depend strongly on the footprint of $E_i$ on the RIS. To investigate the role of the incident beam's footprint, let us consider a Gaussian beam illuminating the RIS from the angle $\theta_i$. The incident $E$-field at the RIS surface can be written as \cite{Stratidakis2022}
\begin{equation}
    E_i(x,y) = E_0 \exp\left(-\frac{x^2 \cos^2\theta_i+y^2}{w^2_\mathrm{RIS}} \right)e^{-jk_i x},
    \label{Eq:EqEincGB}
\end{equation}
where $w_\mathrm{RIS}$ is the beam's radius at normal incidence ($\theta_i=0$). Using Eqs.\@(\ref{Eq:EqErINT}), \@(\ref{Eq:EqTRISc}), and \@(\ref{Eq:EqEincGB}) we find that
\begin{equation}
    \widetilde{E}_r = \Gamma_0 \widetilde{E}_i(k_x+k_i-k_r,k_y) R(k_x,k_y),  
    \label{Eq:EqErflGB}
\end{equation}

where

\begin{equation}
    \widetilde{E}_i(k_x,k_y) = E_0 \frac{\pi w_\mathrm{RIS}^2}{\cos{\theta_i}} e^{-\frac{w_\mathrm{RIS}^2}{4} \left(\frac{(k_x-k_i)^2}{\cos^2{\theta_i}}+k_y^2\right)}
    \label{Eq:EqEincGBk}
\end{equation}
is the Fourier transform of \eqref{Eq:EqEincGB} and
\begin{equation}
    R(k_x,k_y) = \frac{\mathrm{erf}(q_x)+\mathrm{erf}(q_x^*)}{2}\frac{\mathrm{erf}(q_y)+\mathrm{erf}(q_y^*)}{2},
\end{equation}
where erf$\left(\cdot\right)$ is the error function and $q_x,q_y$ are given by
\begin{subequations}
    \begin{gather}
        q_x = \frac{1}{2} \left( \frac{L_x\cos{\theta_i}}{w_\mathrm{RIS}} + j \frac{(k_x-k_r) w_\mathrm{RIS}}{\cos{\theta_i}}\right), \\
        q_y = \frac{1}{2} \left( \frac{L_y}{w_\mathrm{RIS}} + j k_y w_\mathrm{RIS}\right).
    \end{gather}
\end{subequations}
%
%
%
\begin{figure}[t!]
\centering
    \includegraphics[width=\linewidth]{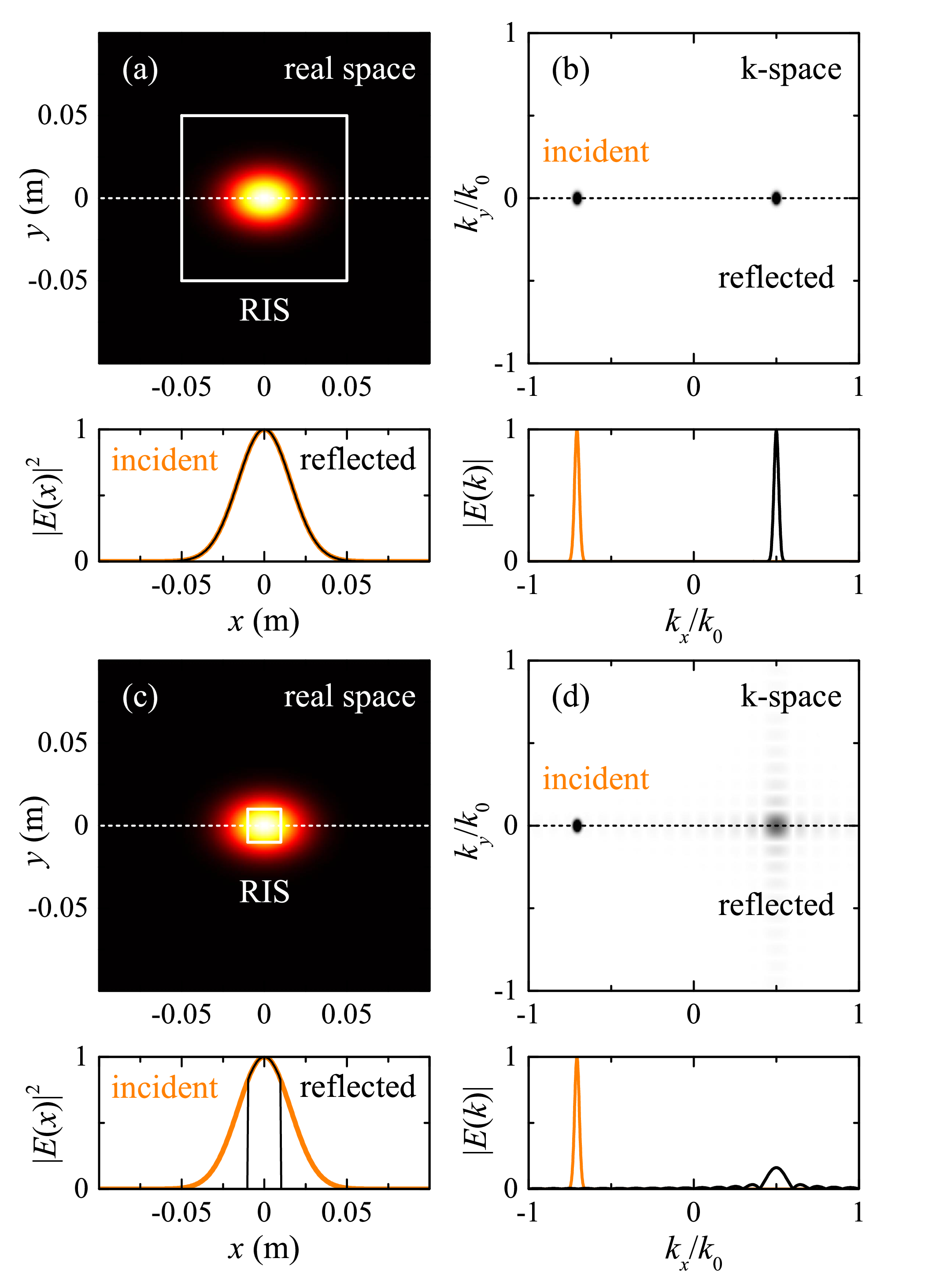} 
	\caption{Impact of the finite RIS size on the incident beam and reflection. (a),(b) $L_x,L_y>w_\mathrm{RIS}$ and (c),(d) $L_x,L_y<w_\mathrm{RIS}$. The footprints of the incident and reflected beam at the RIS ($z=0$) are shown in (a),(c) in real space, where the white square marks the boundary of the RIS. In (b),(d) the same beam footprints are shown in the $k$-space. In (a),(b) the beam footprint is much smaller than the RIS area, thus, the reflected $k$-content is dominated by the incident $k$-content. In (c),(d) the beam footprint is much larger than the RIS area, thus, the reflected $k$-content is dominated by the RIS transfer function. The horizontal dashed lines mark the cross-sections shown below each panel.}
    	\label{fig:fig03}
\end{figure}
\indent In \@(\ref{Eq:EqErflGB}), the $k$-content of the reflected beam is essentially decomposed into the individual contributions from the incident beam ($\widetilde{E}_i$ term) and the RIS transfer function ($R$ term). The balance between the two contributions in the total $k$-content of the reflected beam is determined by the beam footprint relative to the RIS size, as expressed by the functional form of $R(k_x,k_y)$. For example, for $w_\mathrm{RIS} \ll L_x, L_y$, $R(k_x,k_y) \rightarrow 1$ and the effect of the RIS boundary is eliminated. Essentially, the entire beam is captured by the RIS and the resulting beam is the same as in the case of an infinitely large RIS (we refer to this case as \textit{partial illumination}). In this case, the RIS imposes a simple $k$-shift without modifying the incident $k$-content and, therefore, the reflected $k$-content is determined entirely by the incident beam. This is demonstrated in Fig.\,\ref{fig:fig03}(a), where we plot the footprint of a pencil beam of radius $w_\mathrm{RIS}=0.02\,\mathrm{m}$, incident on a RIS of size $L_x=L_y=0.1\,\mathrm{m}$ at angle $\theta_i=45^\circ$, operating at $150\,\mathrm{GHz}$ with $\Gamma_0=1$ (the white square marks the boundary of the RIS). The $k$-content of the incident beam shown in Fig.\,\ref{fig:fig03}(b) is simply shifted by $k_r-k_i$, as in the case of infinitely large RIS. On the other hand, for a beam footprint larger than the RIS area, the RIS size plays the dominant role; in this case, the RIS can be considered as an effective radiating aperture of size $L_x \times L_y$, which modifies the incident $k$-content according to the sinc-like profile of \@(\ref{Eq:EqTRISc}) (we refer to this case as \textit{full illumination}). This is demonstrated in Fig.\,\ref{fig:fig03}(c), where the same beam as in Fig.\,\ref{fig:fig03}(a) illuminates a smaller RIS of size $L_x=L_y=0.01\,\mathrm{m}$. Clearly, the sinc-like form of $\widetilde{T}_\mathrm{RIS}$ (see \eqref{Eq:EqTRISc}) now dominates the $k$-content of the reflected beam, as shown in Fig.\,\ref{fig:fig03}(d). This implies that the reflected beam has a substantially larger $k$-content than the incident beam. This is a typical situation in LOS propagation scenarios where the RIS is in the far-field of the transmitter so the incident wave is approximately plane over the RIS \cite{Bjornson2020b}.
\section{Far-field power distribution}
\noindent The $k$-content of the reflected wave in the RIS plane also provides useful information about its far-field distribution. It is well known from Fourier optics that a wave at the image plane corresponds to the Fourier transform of its spatial distribution in the object plane \cite{Goodman2005, Droulias2022}, i.e., propagation in free space is equivalent to Fourier-transforming the input wave. Here, the object plane is considered at the RIS, while the image plane is at the location of the receiver. \\
\indent As an example, let us consider a rectangular RIS of size $L_x \times L_y$. The RIS is illuminated by a TE-polarized beam of total power $P_t$, emitted from the access point (AP), which is located at distance $d_\mathrm{AP}$ from the RIS and is equipped with a directional antenna that has gain $G_t$. The user equipment (UE) is located at the point defined by the observation vector $\textbf{r} = x\hat{\textbf{x}}+y\hat{\textbf{y}}+z\hat{\textbf{z}}$ (or, in spherical coordinates, $\textbf{r} = r\sin{\theta}\cos{\varphi}\hat{\textbf{x}} + r\sin{\theta}\sin{\varphi}\hat{\textbf{y}} + r\cos{\theta}\hat{\textbf{z}}$). \\
\indent Following the derivation in \cite{Droulias2023} for surfaces that are significantly larger than $\lambda$ (as is the case in the intended practical deployment situations), we find here that the power received by a UE that is located at the observation point ($r, \theta, \varphi$) in the far-field of the RIS, i.e., at $|r|>$2\,max$(L_x^2,L_y^2)/\lambda$, is given by (see Appendix \ref{AppendixC} for details)

\begin{equation}
    P_r = A_r \frac{k^2}{2Z_0} \Theta(\theta,\varphi) \frac{\left|\widetilde{E}_r(k_x,k_y)\right|^2}{(4\pi r)^2},
    \label{Eq:EqPr}
\end{equation}
where
\begin{align}
    \begin{split}
        \Theta(\theta,\varphi) = \sin^2{\varphi}(1+\cos{\theta}\cos{\theta_r})^2\\
        +\cos^2{\varphi}(\cos{\theta}+\cos{\theta_r})^2
    \end{split}
\end{align}
and $\widetilde{E}_r(k_x,k_y)$ accounts for the $k$-content of the Gaussian beam that is given by \eqref{Eq:EqErflGB}, with $k_x = k_0\sin{\theta}\cos{\varphi}$, $k_y = k_0\sin{\theta}\sin{\varphi}$. \\

%
%
\begin{figure}[t!]
\centering
    \includegraphics[width=\linewidth]{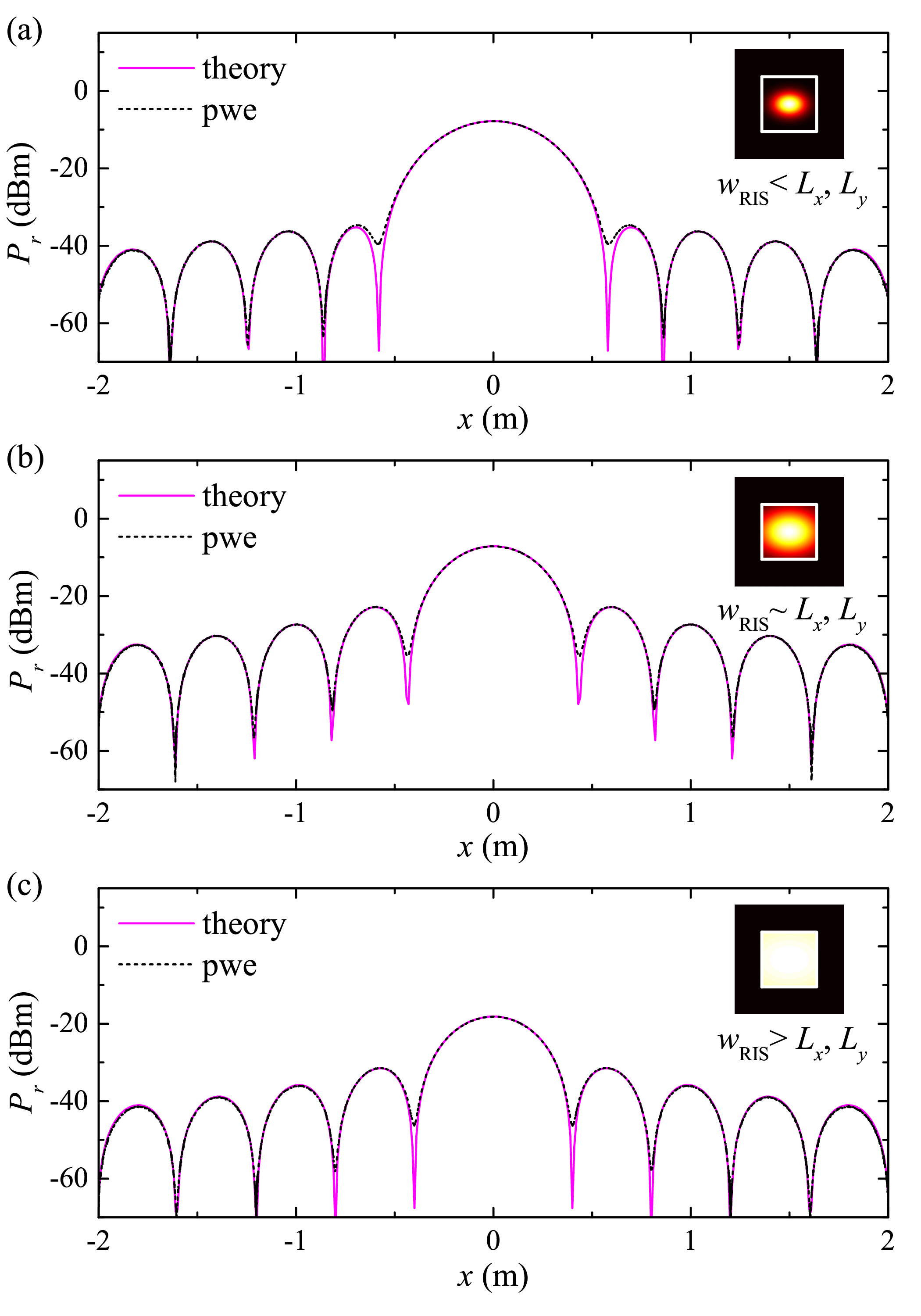}  
	\caption{The analytical far-field distribution of the received power (calculated at $z = 20\,\mathrm{m}$ from the RIS) and the numerical validation using the plane wave expansion (pwe). The AP has 40 dB gain and illuminates the RIS at angle $\theta_i=45^\circ$ and the RIS steers the beam towards $\theta_r=0^\circ$. The AP beam footprint on the RIS (shown in the inset) is controlled via $d_\mathrm{AP}$, the distance between the AP and RIS. (a) $d_\mathrm{AP}=1\,\mathrm{m}$, (b) $d_\mathrm{AP}=2\,\mathrm{m}$, and (c) $d_\mathrm{AP}=10\,\mathrm{m}$. The RIS is operating at $150\,\mathrm{GHz}$ and consists of $250\times 250$ elements with periodicity $l_x=l_y=\lambda/5$ (size $10\,\mathrm{cm}\times 10\,\mathrm{cm}$).}
    	\label{fig:fig04}
\end{figure}
\indent In Fig.\,\ref{fig:fig04} we use \eqref{Eq:EqPr} to calculate the far-field distribution of the received power for a RIS that steers the incident beam towards $\theta_r=0^\circ$ (solid magenta lines). The receiver is located on the $z$-axis at the distance $z=20\,\mathrm{m}$ from the RIS. To validate the analytical expression we use the plane wave expansion to numerically propagate the $E$-field from the RIS (as given by \eqref{Eq:EqEincGB}) to the far-field (see Appendix \ref{AppendixD} for details). That is, to find the field at the receiver, we numerically perform the following operation: 
\begin{equation}
    E_\mathrm{num}(x,y,z) =\mathcal{FT}^{-1}\{\,\mathcal{FT}\,[E_r(x,y)]e^{jk_zz}\},
    \label{Eq:EqEnum}    
\end{equation}
where \(\mathcal{FT}\) denotes the Fourier transform and \(\mathcal{FT}\)\,$^{-1}$ its inverse, $E_r(x,y)=\Gamma(x,y)E_i(x,y)$ is the reflected footprint at $z=0$ and $k_z=\sqrt{k^2_0-k^2_x-k^2_y}$ is the transverse wavenumber. Then, we use the numerically calculated $E$-field, $E_\mathrm{num}$, to find the received power $P_{r,num}=A_r|E_\mathrm{num}|^2/2Z_0$ (dashed black lines). The RIS is operating at 150 GHz and consists of $250\times 250$ elements with periodicity $l_x=l_y=\lambda/5$ (size $10\,\mathrm{cm}\times 10\,\mathrm{cm}$). The AP has 40 dB antenna gain and illuminates the RIS from the angle $\theta_i=45^\circ$ with a TE-polarized Gaussian beam, the footprint of which on the RIS is controlled via $d_\mathrm{AP}$, the distance of the AP from the RIS. In Fig.\,\ref{fig:fig04}(a) $d_\mathrm{AP}=1\,\mathrm{m}$ and the beam footprint, which is smaller than the RIS extent, is captured entirely by the RIS. In this case, the RIS boundaries do not interfere with the beam, which is scattered as if the RIS practically had an effectively infinite extent. In Fig.\,\ref{fig:fig04}(b) $d_\mathrm{AP}=2\,\mathrm{m}$ and the RIS partially truncates the beam, as shown in the inset. Last, in Fig.\,\ref{fig:fig04}(c) $d_\mathrm{AP}=10\,\mathrm{m}$ and the relatively wide beam practically illuminates the entire RIS panel almost uniformly, as shown in the inset.
\section{Advanced RIS operations}
\subsection{Far-field beam shape engineering}

%
%
\begin{figure}[t!]
\centering
    \includegraphics[width=\linewidth]{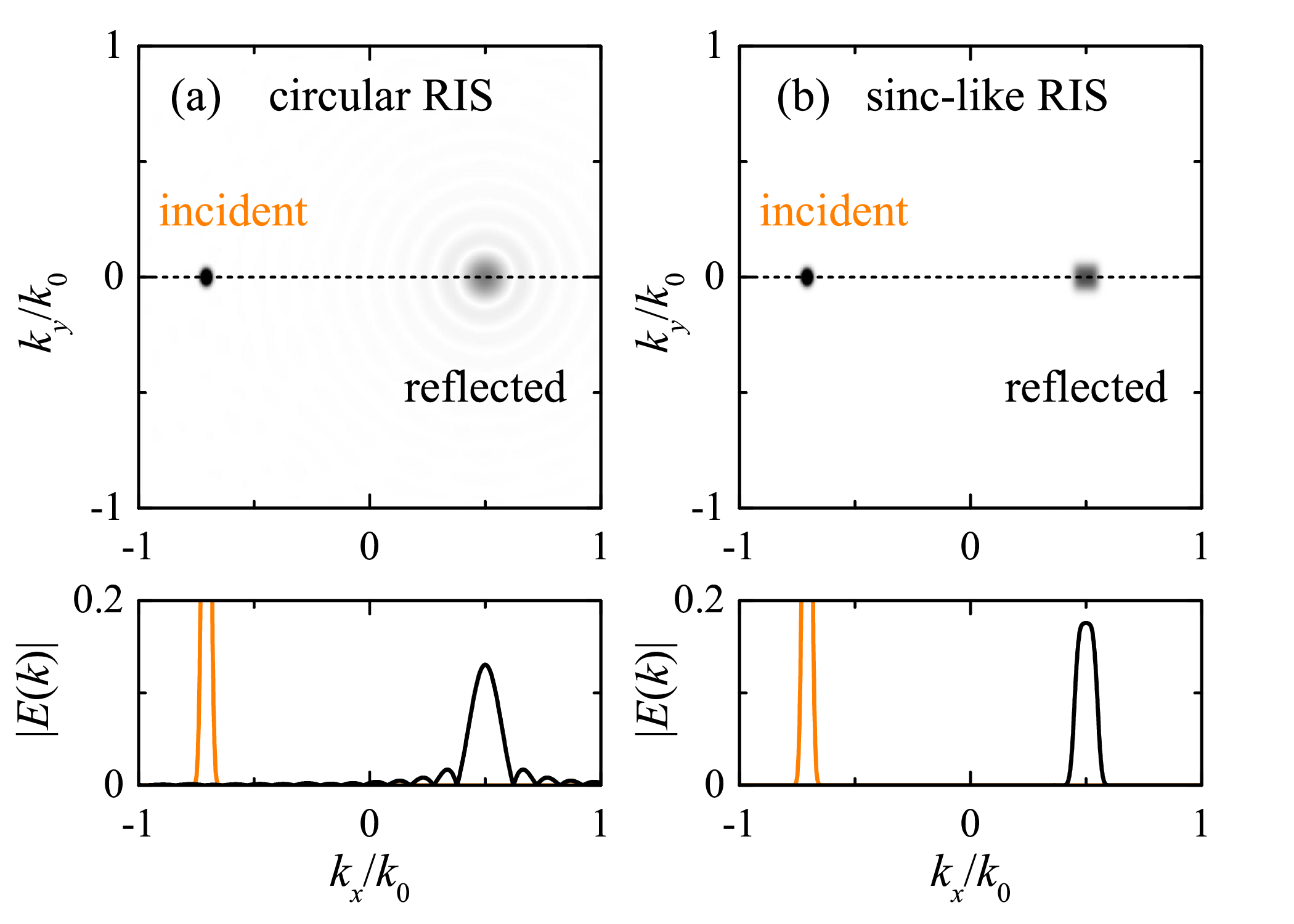} 
	\caption{Engineering the far-field via the RIS shape. $k$-content of the reflected wave for fully illuminated RIS of (a) circular shape and (b) sinc-like shape. In (b) the side lobes are suppressed in the far-field, via tailoring the RIS shape. The horizontal dashed lines mark the cross-sections shown below each panel; in the bottom row panels, the incident $k$-content has been truncated to emphasize the details of the reflected beams.}
    	\label{fig:fig05}
\end{figure}

\noindent By tailoring the RIS transfer function, it is possible to control the properties of the wave in the far-field beyond the sinc-like pattern achieved with a typical rectangular RIS. This can be achieved simply by changing the shape of the RIS, while maintainging the typical linear phase profile, i.e., by modifying the functional form of $\Gamma_0(x,y)$. For example, under full illumination conditions, a circular RIS may lead to a Bessel-like far-field pattern, as illustrated in Fig.\,\ref{fig:fig05}(a). In general, any arbitrary shape will lead to its Fourier transform in the far-field and, hence, we can reverse engineer this operation to control the desired far-field via appropriate tuning of the RIS's properties. Ideally, we can always perform the inverse Fourier transform of an arbitrary far-field distribution and then search for the RIS configuration that can approximate the desired operation. For example, if we wish to minimize crosstalk we can entirely suppress side lobes by using a sinc-like RIS (e.g. by introducing absorptive regions in the RIS), as shown in Fig.\,\ref{fig:fig05}(b). In general, side lobes can also be suppressed by tapering the incident field with the RIS or by illuminating the RIS with a tapered beam, as in the example of Fig.\,\ref{fig:fig03}(a) for partial illumination conditions.
\subsection{Multi-beam operation}
\noindent Besides single beam formation, the RIS can split the incident beam into multiple reflected beams, to multicast to selected multiple users \cite{Tsang2011}. In this case, each reflected beam is associated with an individual reflection coefficient $\Gamma_n(x,y) = \Gamma_0e^{j(k_{r,n}-k_i)x}$, where $k_{r,n}$ is the wavenumber of the $n^\mathrm{th}$ reflected beam. The RIS transfer function is then written in real space as
\begin{align}
    T_\mathrm{RIS}(x,y,x',y') = \sum_{n=1}^N \frac{w_n}{w_0}\Gamma_n(x,y) \delta(x-x',y-y'),    
    \label{Eq:EqMB_TRIS}
\end{align}
where $w_n$ accounts for the weighted contribution of the $n^\mathrm{th}$ beam to the total field, and $w_0$ is a normalization constant. Using \eqref{Eq:EqMB_TRIS}, the reflected field at $z=0$ is expressed as a linear superposition of the $N$ individual beams as
\begin{align}
    E_r(x,y) = E_i(x,y)\sum_{n=1}^N \frac{w_n}{w_0} \Gamma_n(x,y).    
    \label{Eq:EqMBEr}
\end{align}
\noindent The weights $w_n$ can be chosen arbitrarily to tune the relative contributions among the beams, however the normalization constant $w_0$ must ensure that, for a passive and lossless RIS, the incident power is preserved upon reflection, i.e. that $\iint_{A_\mathrm{RIS}}|E_r(x,y)|^2dxdy = \iint_{A_\mathrm{RIS}}|E_i(x,y)|^2dxdy$, where the integration is performed within $A_\mathrm{RIS}$, the RIS area. This constraint leads to
\begin{align}
w_0=\sqrt{\frac{\iint_{A_\mathrm{RIS}}|E_i(x,y)|^2|\sum_{n=1}^N w_n\Gamma_n(x,y)|^2dxdy}{\iint_{A_\mathrm{RIS}}|E_i(x,y)|^2dxdy}}.
\label{Eq:EqW0}
\end{align}
For full illumination, $|E_i(x,y)|$ is constant throughout the RIS surface and, hence, \eqref{Eq:EqW0} is simplified as
\begin{align}
w_0=\sqrt{\frac{\iint_{A_\mathrm{RIS}}|\sum_{n=1}^N w_n\Gamma_n(x,y)|^2dxdy}{A_\mathrm{RIS}}}.
\label{Eq:EqW0FULL}
\end{align}
A common $w_n$ for all $n$ corresponds to distributing the incident power equally among the $N$ beams. In this case, the prefactor $w_n/w_0$ in \eqref{Eq:EqMB_TRIS} and \eqref{Eq:EqMBEr} becomes
\begin{align}
\frac{w_n}{w_0}=\sqrt{\frac{A_\mathrm{RIS}}{\iint_{A_\mathrm{RIS}}|\sum_{n=1}^N\Gamma_n(x,y)|^2dxdy}}.
\label{Eq:EqW0FULLeq}
\end{align}

%
%
\begin{figure}[t!]
    \centering
    \includegraphics[width=\linewidth]{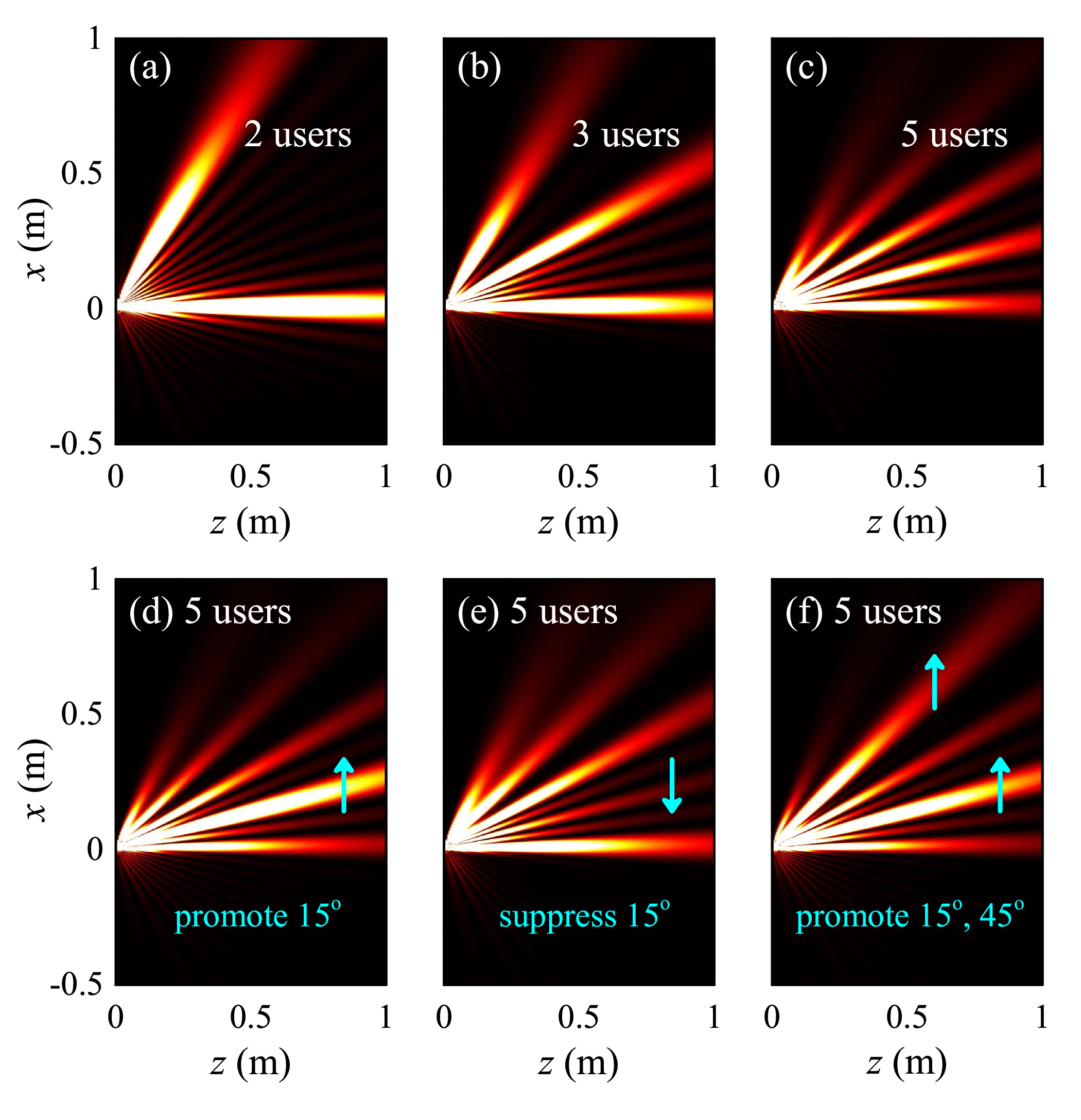}
    \caption{Multiple simultaneously reflected beams, for broadcast to selected multiple users. A fully illuminated RIS of size $2\,\mathrm{cm} \times 2\,\mathrm{cm}$ splits the incident beam into (a) $N=2$, (b) $N=3$, and (c) $N=5$ beams of equal power, along the directions characterized by the angles $\theta_{r,n}=(n-1)\times 60^\circ/(N-1)$, $n=1,2,\dots,N$. (d) Promotion of beam along the direction $\theta_r=15^\circ$ ($N=5$) and (e) suppression of the same beam. (f) Promotion of two beams along the directions $\theta_r=15^\circ,45^\circ$.}
	\label{fig:figMU}
\end{figure}
As an example, in Fig. \ref{fig:figMU} we propagate numerically multiple beams that are simultaneously reflected by a fully illuminated RIS consisting of $50\times 50$ elements with periodicity $l_x=l_y=\lambda/5$ (size $2\,\mathrm{cm}\times 2\,\mathrm{cm}$).
The incident beam is reflected into $N=2$ in Fig. \ref{fig:figMU}(a), $N=3$ in Fig. \ref{fig:figMU}(b), and $N=5$ in Fig. \ref{fig:figMU}(c) beams, which are directed towards the angles $\theta_{r,n}=(n-1)\times 60^\circ/(N-1)$, $n=1,2,\dots,N$. In these examples $w_n=1$ for $n=1,2,\dots,N$, to achieve equal power split among the $N$ beams. Next, we tune the relative weights to either promote or suppress the reflection among certain directions. For example, in Fig. \ref{fig:figMU}(d), we keep $w_2=1$ and select $w_1=w_3=w_4=w_5=0.5$, to promote the beam along the direction $\theta_r=15^\circ$. In Fig. \ref{fig:figMU}(e), we keep $w_1=w_3=w_4=w_5=1$ and select $w_2=0.25$, to suppress the same beam. Last, in Fig. \ref{fig:figMU}(f) we choose $w_1=w_3=w_5=0.5$ and $w_2=w_4=1$, to promote two beams along the directions $\theta_r=15^\circ, 45^\circ$.

\indent In general, any combination of directions and weights is possible for any number of beams, making this scheme ideal for applications including broadcasting to selected multiple users, tracking and positioning algorithms, user grouping for advanced scheduling and multiple access schemes. For example, in a Time Division Multiple Access (TDMA) scheme, the incident beam is directed towards different directions at different time slots, to successively serve each user. Similarly, in a Space Division Multiple Access (SDMA) scheme, different areas of the RIS can be devoted to simultaneously send multiple signals towards multiple users.
\subsection{Interference suppression}
\noindent In the examples described so far, the wave transformation typically requires that a single wave is incident on the RIS. In realistic situations, however, it is expected that several waves simultaneously impinge on the RIS. For example, if we have a desired receiver location and one desired incident plane wave to the RIS, there may be additional interfering plane waves. In this case, how can we suppress and even entirely eliminate the reflections of those interfering plane waves at the desired receiver? One approach to answering this question is to apply a filter in the $k$-space that preserves only $k_i$, the incident wave direction that we desire to steer, while eliminating all other incident $k$'s. To achieve this, we need to apply a spatial bandpass filter function centered at $k_i$, with bandwidth tailored to capture the entire $k$-content of the desired incident wave and eliminate any other wave outside the desired $k$-window. \\
\indent Using the filter function $\exp\left(-(k_x-k_i)^2/2 k_F^2\right)$, where $2\sqrt{2\ln{2}}k_F$ is its Full Width at Half Maximum (FWHM), the RIS transfer function becomes
\begin{equation}
        \widetilde{T}_\mathrm{RIS}=\Gamma_0 e^{-\frac{(k_x'-k_i)^2}{2 k_F^2}} \delta(k_x-k_x'+k_i-k_r,k_y-k_y').
    \label{Eq:EqTRISfilterK}
\end{equation}
\indent In real space, this operation leads to a transfer function of the form (see Appendix \ref{AppendixE} for details):
\begin{equation}
        T_\mathrm{RIS}=\Gamma_0 \frac{k_F}{\sqrt{2\pi}} \exp \left(-\frac{k_F^2}{2}(x-x')^2 + j(k_r x - k_i x')\right).
    \label{Eq:EqTRISfilterX}
\end{equation}
Then, if the wave incident on the RIS is a mixture of the desired wave and a wave impinging at the angle $k_n$ with amplitude $E_n$, $E_i = E_0e^{j k_i x} + E_n e^{j k_n x}$, application of \eqref{Eq:EqTRISfilterX} leads to a reflected wave of the form
\begin{equation}
    E_r = \Gamma_0 e^{j k_r x} \left(E_0 + E_n e^{-j(k_i-k_n)x} e^{-\frac{(k_i-k_n)^2}{2k_F^2}} \right).
\end{equation}
Clearly, for $|k_i-k_n|$ larger than $\approx3k_F$, the unwanted wave is practically eliminated and the reflected wave is entirely determined by the desired incident wave characterized by $k_i$. \\
%

%
%
\begin{figure}[t!]
    \centering
    \includegraphics[width=\linewidth]{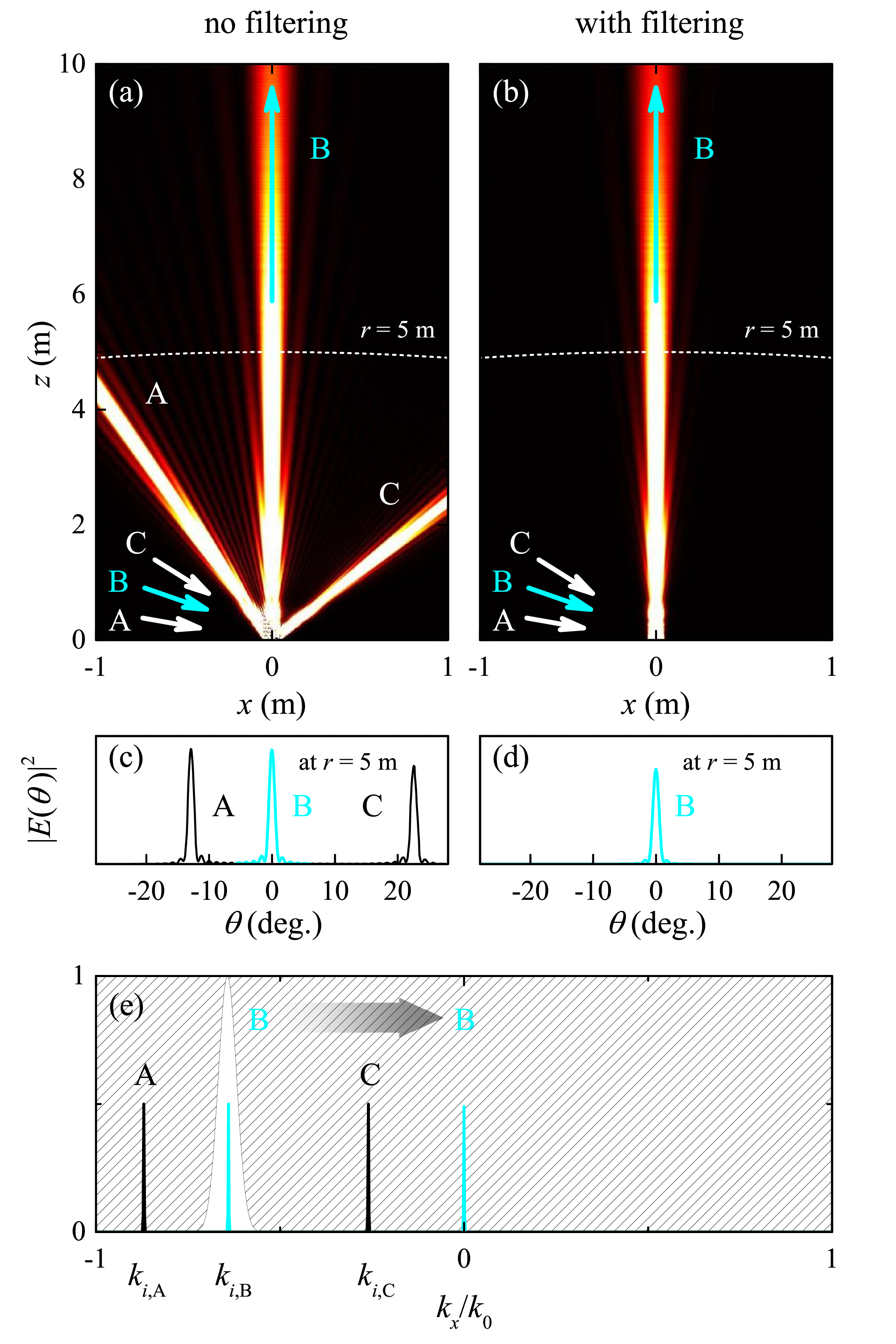}
    \caption{Example of interference suppression by spatial bandpass filtering. The RIS, located at $z=0$, is designed to steer the incident wave B from $\theta_i=40^\circ$ towards $\theta_r=0^\circ$. The undesired waves A and C are (a) steered in the absence of a filter and (b) suppressed when spatial filtering is applied. The white dashed line marks the distance $r=5\,\mathrm{m}$, at which beam cross-sections as a function of the observation angle $\theta$ are shown in (c) and (d), without and with filtering, respectively. (e) Operation of the spatial bandpass filtering operation in the $k$-space. The filter function is a Gaussian distribution, centered at $k=k_{i,B}$ with $k_F=0.025k_0$.}
	\label{fig:fig07}
\end{figure}

\indent As an example, in Fig.\,\ref{fig:fig07}, we demonstrate interference suppression with a RIS characterized by the transfer function \eqref{Eq:EqTRISfilterK}. The RIS consists of $100\times 100$ elements with periodicity $l_x=l_y=\lambda/2$ (size $10\,\mathrm{cm}\times 10\,\mathrm{cm}$) and is designed to steer a single wave from $\theta_i=40^\circ, \varphi_i=\pi$ to $\theta_r=0^\circ$. Besides the desired wave that impinges at the expected angle $\theta_{i,B}=40^\circ$, two unwanted waves arrive at angles $\theta_{i,A}=60^\circ$ and $\theta_{i,C}=15^\circ$, all traveling towards the $+x$ direction ($\varphi_{i,A}=\varphi_{i,B}=\varphi_{i,C}=\pi$). When the RIS performs plain steering without filtering, all three beams undergo steering towards individual directions, characterized by $\theta_{r,A}=-12.9^\circ$, $\theta_{r,B}=0^\circ$ and $\theta_{r,C}=22.6^\circ$, as shown in Fig.\,\ref{fig:fig07}(a). With filtering, the unwanted beams A and C are eliminated and only the desired beam B is steered as intended, as shown in Fig.\,\ref{fig:fig07}(b). Cross-sections of the beams at distance $r=5\,\mathrm{m}$, as a function of the observation angle $\theta$, are shown in Fig.\,\ref{fig:fig07}(c) and Fig.\,\ref{fig:fig07}(d), for no filtering and filtering, respectively. Due to the bandpass nature of the filtering operation, the filtered beam B has slightly reduced power. The operation of spatial filtering is also schematically shown in Fig.\,\ref{fig:fig07}(e). The Gaussian filter of extent $k_F=0.025 k_0$ is centered at $k=k_{i,B}$, in order to eliminate the waves with $k=k_{i,A},k=k_{i,C}$, while shifting $k_{i,B}$ to $k_{r,B}=0$.
\section{RIS phase optimization}
\begin{figure}[t!]
\centering
    \includegraphics[width=\linewidth]{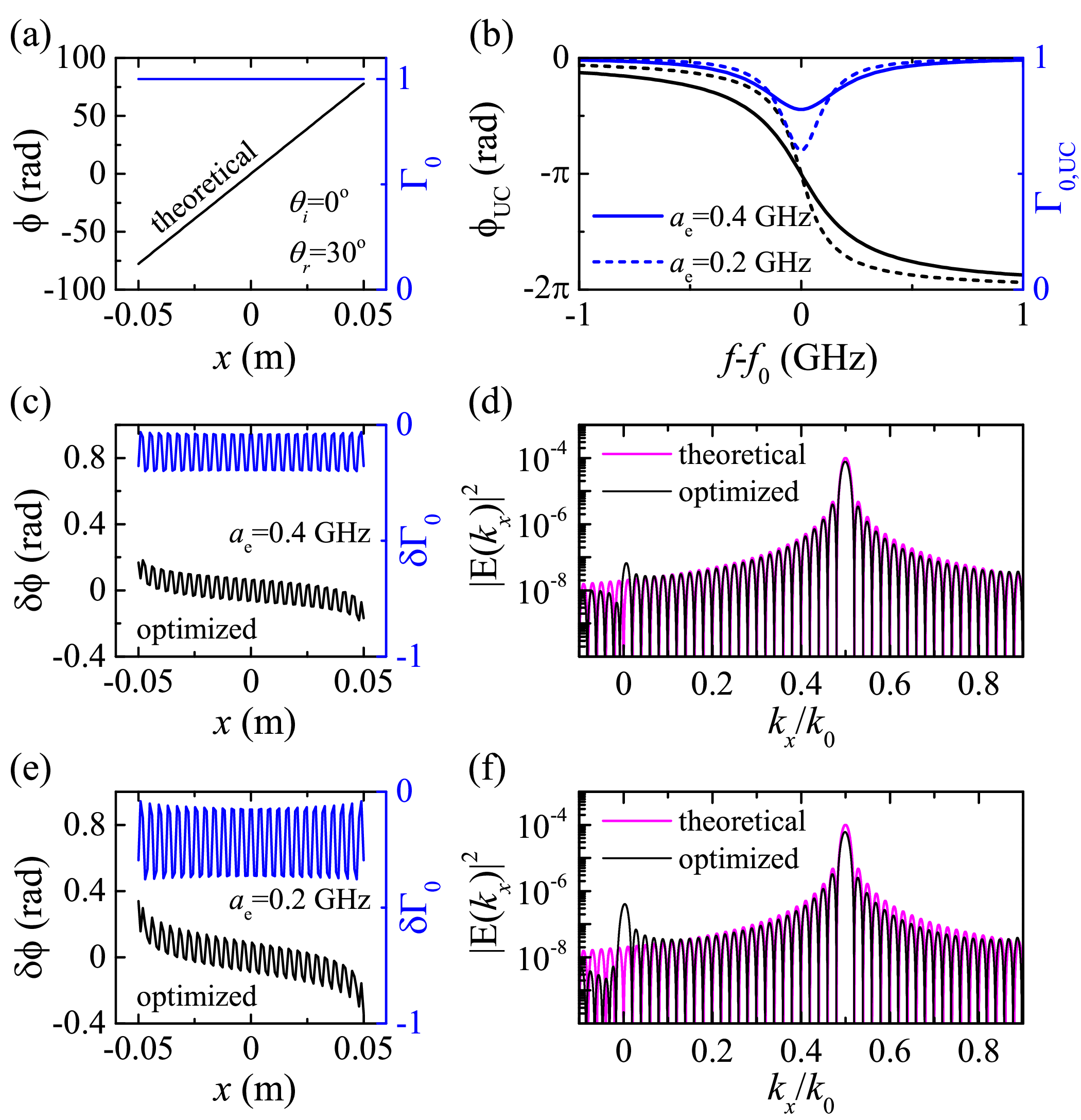}
	\caption{RIS phase optimization. (a) Phase and amplitude of theoretical RIS reflection coefficient distribution $\Gamma(x)$, required for steering a beam from $\theta_i=0^\circ$ towards $\theta_r=30^\circ$. (b) Phase and amplitude of RIS element reflection coefficient $\Gamma_\mathrm{UC}$, as a function of the element frequency detuning $f-f_0$, for RIS elements with Lorentzian response. (c),(e) Optimized RIS reflection coefficient distribution $\Gamma_\mathrm{opt}(x)$, for the example element reflection coefficient presented in (b). (d),(f) Theoretical $k$-content of RIS footprint and optimized $k$-content, for the respective $\Gamma_\mathrm{opt}(x)$ in (c),(e).}
    	\label{fig:fig08}
\end{figure}

\noindent In the demonstrated examples, the RIS reflection coefficient has tunable phase $\phi$ within the ($-\pi,\pi$) range and uniform amplitude across the entire tunable range ($\Gamma_0=1$ for lossless RIS). While this condition is met in several experimental works \cite{Yang2016,Zhang2018,Dai2020}, in general it could be rather restrictive. The reason is that, because the RIS reflection coefficient is usually tuned via the resonances of the constituent scatterers, tuning of the RIS elements is typically associated with changes in both the phase and the amplitude of $\Gamma$ \cite{Bjornson2021, DeRosny2021}. As a result, tuning the RIS elements to the desired phase leads to non-uniform amplitude, with implications on the RIS performance. \\
\indent Taking advantage of the fact that the far-field of the reflected beam is determined entirely by the $k$-content of its footprint on the RIS, $\widetilde{E}_r(k_x,k_y)$, we can use our framework to optimize the RIS phase shifts, so that the corresponding RIS transfer function $T_\mathrm{RIS}$ leads to an optimized footprint, which is as close as possible to the theoretical. The theoretical footprint is associated with the RIS reflection coefficient given by \eqref{Eq:EqREFLCOEFF}, with $\Gamma_0(x,y)=1\equiv \Gamma_{0,\mathrm{theory}}$ and $\phi(x,y) = k_0(\sin{\theta_r}-\sin{\theta_i})x \equiv \phi_\mathrm{theory}$; therefore, we denote this footprint as $\widetilde{E}_r(k_x,k_y;\Gamma_\mathrm{theory})$. We denote as $\widetilde{E}_r(k_x,k_y;\Gamma)$ the footprint to be optimized, in which the RIS reflection coefficient is given by \eqref{Eq:EqREFLCOEFF}, with $\Gamma_0(x,y)$ and $\phi(x,y)$ provided by the specific design of the unit cell (UC) reflection coefficient $\Gamma_\mathrm{UC}(f)=\Gamma_{0,\mathrm{UC}}(f)e^{j\phi_\mathrm{UC}(f)}$, where $\Gamma_{0,\mathrm{UC}}=|\Gamma_\mathrm{UC}|$ is the UC reflection amplitude, $\phi_\mathrm{UC}$ is the UC reflection phase, and $f$ is the operation frequency. As a result, the amplitude $\Gamma_0(x,y)$ and phase $\phi(x,y)$ at any $(x,y)$ location on the RIS are not independent anymore. The optimization problem is then formulated as
\begin{subequations}
    \begin{align}
        &\min_{\Gamma}|\widetilde{E}_r(k_x,k_y;\Gamma)-\widetilde{E}_r(k_x,k_y;\Gamma_\mathrm{theory})| \\
        \mathrm{s.t.}\,\; &\Gamma(x,y) \in \Gamma_{0,\mathrm{UC}}e^{j\phi_\mathrm{UC}} \\
        &\Gamma_{0,\mathrm{UC}} = \Gamma_{0,\mathrm{UC}}(\phi_\mathrm{UC}).
    \end{align}
    \label{Eq:EqOPT}
\end{subequations}
\noindent Simply put, the footprint is optimized under the constraint that the RIS element reflection coefficient is restricted to the values provided by the UC reflection coefficient [\eqref{Eq:EqOPT}(b)], for which the available phases and amplitudes are not independent, but are associated via \eqref{Eq:EqOPT}(c). \\
\indent As an example, we consider a $10\,\mathrm{cm}\times 10\,\mathrm{cm}$ RIS  ($100\times 100$ elements with periodicity $l_x=l_y=\lambda/2$) operating at $f=150\,\mathrm{GHz}$. The RIS steers waves from $\theta_i=0^\circ$ towards $\theta_r=30^\circ$, and the theoretical reflection coefficient required for this operation is shown in Fig.\,\ref{fig:fig08}(a). For the RIS elements we consider a typical Lorentzian response \cite{Abeywickrama2020, Bjornson2021, DeRosny2021, Abbas2023, Hassouna2024}, which we use to analytically derive a theoretical model for $\Gamma_\mathrm{UC}$ (see Appendix \ref{AppendixF} for details on the derivation). The amplitude and phase of each RIS element is given analytically by
\begin{equation}
    \Gamma_{0,\mathrm{UC}}(f)=\sqrt{1-\frac{4a_e\gamma_ef^2}{(\gamma_e-a_e)^2f^2 + (f^2-f_0^2)^2}}
    \label{Eq:GammaUCabs}
\end{equation}
and
\begin{equation}
    \phi_\mathrm{UC}(f)=\arctan{\left(\frac{2a_ef(f^2-f_0^2)}{(\gamma_e^2-a_e^2)f^2 + (f^2-f_0^2)^2}\right)},
    \label{Eq:GammaUCarg}
\end{equation}
respectively, where $f$ is the operation frequency, $f_0$ the resonance frequency of the RIS element, $\gamma_e$ the damping term of the element response and $a_e$ a parameter that tunes the resonance strength. In our example $f_0$ is varied to detune the RIS element response, in order to gain access to the ($-\pi,\pi$) phase span. The RIS elements are characterized by $\gamma_e=0.05\,\mathrm{GHz}$, and in Fig.\,\ref{fig:fig08}(b) we plot \eqref{Eq:GammaUCabs} and \eqref{Eq:GammaUCarg} for two different values of $a_e$. To solve the optimization problem \eqref{Eq:EqOPT} we express the reflection coefficient to be determined in terms of the theoretical reflection coefficient, as $\Gamma_0 = \Gamma_{0,\mathrm{theory}}+\delta\Gamma_0$, $\phi = \phi_{\mathrm{theory}}+\delta\phi$, where $\delta\Gamma_0$, $\delta\phi$ are the detunings from the theoretical values. Using the optimization procedure outlined in Appendix \ref{AppendixG} for $a_e=0.4\,\mathrm{GHz}$, we find the optimized reflection coefficient shown in Fig.\,\ref{fig:fig08}(c). In Fig.\,\ref{fig:fig08}(d) we compare the $k$-content of the target (magenta line) footprint with that obtained from the optimization scheme (black line). These are essentially cross-sections of the reflected beam's $k$-content at $z=0$. The peak at $k_x/k_0=0.5(=\sin{30^\circ})$ indicates beam formation along the desired direction $\theta_r$, and the optimized beam has slightly lower power from the theoretical, due to $\Gamma_0<1$. For $a_e=0.2\,\mathrm{GHz}$ the RIS reflectivity becomes sharper, leading to larger variations in the optimized reflection coefficient, as shown in Fig.\,\ref{fig:fig08}(e). Due to larger amplitude variations, the integrated power on the RIS is now lower, as is also evident in the corresponding footprint, shown in Fig.\,\ref{fig:fig08}(f). To verify the beam formation, we also numerically propagate the footprints in Fig.\,\ref{fig:fig08}(d),(f) (not shown here). As expected, when the beams evolve into their far-field, their spatial distribution acquires the sinc-like form predicted by their $k$-content, and their main lobe propagates along the theoretically predicted direction.
%

%
%
\indent The above framework can be also applied to experimental data, using \eqref{Eq:GammaUCabs}, \eqref{Eq:GammaUCarg} to fit experimental sets of RIS element reflectivities. Importantly, it can be used to optimize RIS implementations beyond the specific Lorentzian response considered here. In this case, the amplitude and phase of the RIS element reflection coefficient that are associated in a general way, can be modeled by an appropriate fitting function to replace \eqref{Eq:GammaUCabs}, \eqref{Eq:GammaUCarg} in \eqref{Eq:EqOPT}(b)-(d).
\section{Wavefront engineering beyond conventional beamforming}
\noindent Beams generated by small-aperture beamformers quickly enter the far-field, thereafter monotonically spreading with decreasing peak power. However, with wireless communications shifting to higher frequencies, radiating elements become electrically larger, and the transition from the near- to the far-field moves to larger distances, offering new opportunities for taking advantage of the beam's features in its near-field. With electrically large surfaces, the beam's wavefront can be tailored to acquire curvature beyond the typical far-field planar form achieved with conventional beamforming, producing beams with exotic shapes and propagation characteristics \cite{Stratidakis2024, Droulias2024}. \\
\indent In the RIS transfer function framework it is straightforward to extend the RIS operation to account for generalized beams with advanced near-field properties, beyond conventional beamforming. The RIS transfer function essentially acts on the incident beam to first flatten its wavefront [term $\exp(-jk_ix)$] and to subsequently introduce a wavefront tilt towards the steering direction [term $\exp(+jk_rx)$]. If, instead, at the latter step, the RIS transfer function introduces a non-planar wavefront, beams with advanced features can be generated. For example, let us generalize $T_\mathrm{RIS}$ in real space as
\begin{equation}
    T_\mathrm{RIS}(x,y,x',y') = \Gamma_0(x,y)e^{-jk_i x}e^{ja \rho^{\gamma}}\delta(x-x',y-y'), 
    \label{Eq:Eq16}
\end{equation}
to account for radially symmetric beams with power-law wavefronts, where $\rho=\sqrt{x^2+y^2}$ is the radial distance on the RIS, and the parameters $a, \gamma$ shape the reflected wavefront according to the desired beam dynamics. In this formulation, beamsteering along the $x-$axis corresponds to substituting $\rho\rightarrow x$ and setting $a=k_r, \gamma=1$. We can now use \eqref{Eq:Eq16} to generate beams with advanced features.
%

%
%
\begin{figure}[t!]
\centering
    \includegraphics[width=\linewidth]{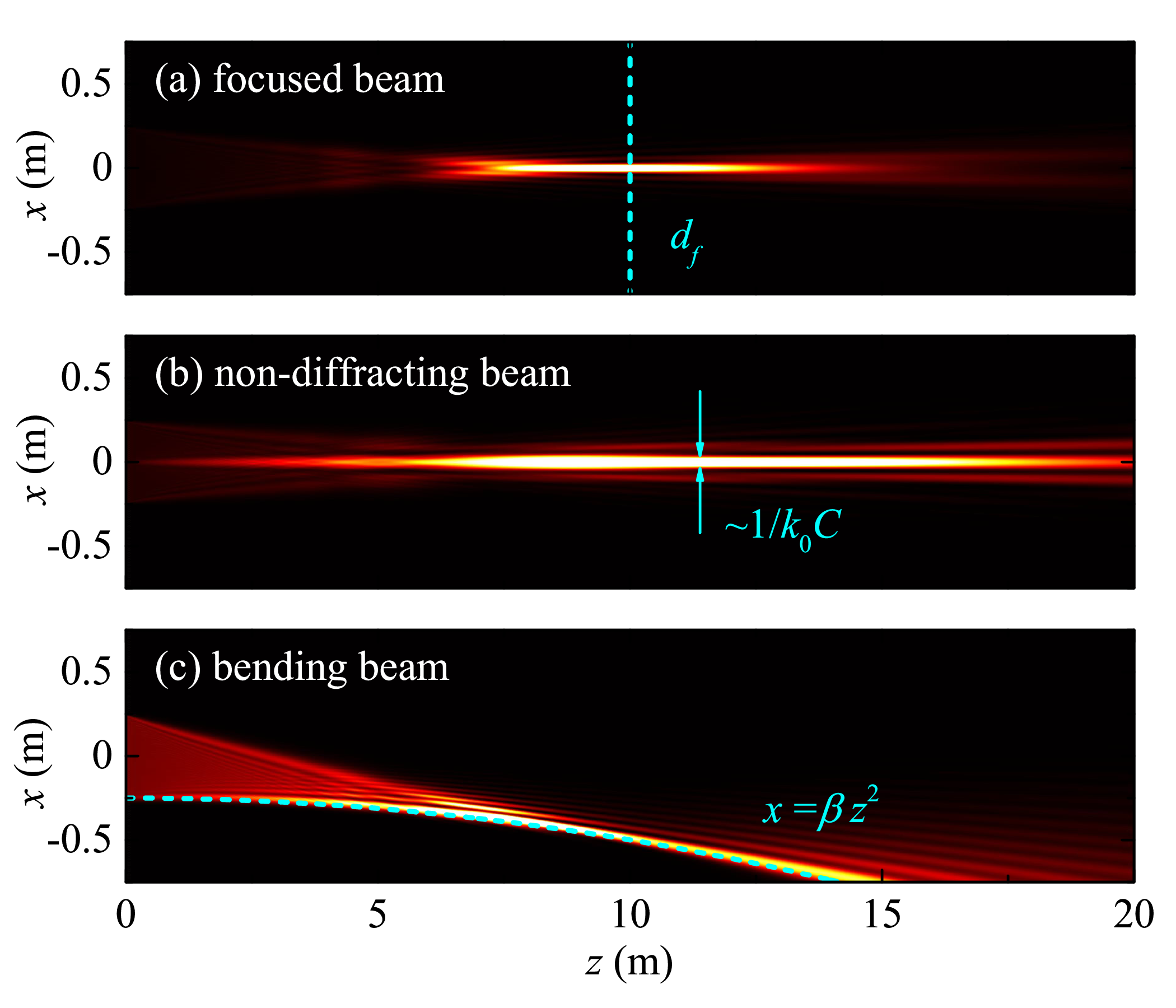}
	\caption{Near-field beam engineering. (a) Beamfocusing at distance $d_f=10\,\mathrm{m}$ from the RIS, marked with the dashed line. (b) Non-diffracting beam with $C=0.0125$, propagating with practically constant main lobe width, for the entire range shown. (c) Bending beam with $\beta=0.0025\,\mathrm{m}^{-1}$, with its main lobe following the parabolic trajectory marked with the dashed line. In all exampled the RIS has size $50\,\mathrm{cm}\times 50\,\mathrm{cm}$ and is fully illuminated ($f=150\,\mathrm{GHz}$).}
    	\label{fig:fig09}
\end{figure}
\indent First, let us consider beams that are capable of focusing. Focused beams offer increased received power in small areas, and are ideal for future applications, including energy efficient communications, wireless power transfer, tracking and localization \cite{Droulias2024}. It is well-known from physical optics that such beams require a parabolic phase profile, i.e.
\begin{subequations}
    \begin{align}
        &a=-k_0/2d_f,\\
        &\gamma=2,
    \end{align}
    \label{Eq:EqADV2}
\end{subequations}
\noindent where $d_f$ is the focal distance. As an example, in Fig.\,\ref{fig:fig09}(a), a large, fully illuminated RIS is utilized to focus the incident beam at focal distance $d_f=10\,\mathrm{m}$. The RIS operates at $150\,\mathrm{GHz}$ and consists of $500\times 500$ elements with periodicity $l_x=l_y=\lambda/2$ (size $50\,\mathrm{cm}\times 50\,\mathrm{cm}$). Note that the Fraunhofer distance, which marks the near-to-far-field transition, is $250\,\mathrm{m}$ in this example, i.e. beamfocusing is formed deep within the near-field. We refer the interested reader to our previous work \cite{Droulias2024}, where we use \eqref{Eq:Eq16} for beamfocusing to analyze several aspects related to the features and efficiency of beamfocusing, in the context of near-field communications.

\indent Next, we use different $\gamma$'s to explore further sophisticated functionalities. For example, for  
\begin{subequations}
    \begin{align}
        &a=-k_0 C,\\
        &\gamma=1,
    \end{align}
    \label{Eq:EqADV3}
\end{subequations}
the reflected beam evolves into a non-diffracting Bessel-beam, i.e. into a beam that propagates with invariant transverse profile \cite{Xiang2020, Goutsoulas2020, Droulias2023b}. $C$ is a parameter that controls the phase oscillations at the input plane and determines the width of its main lobe, as $\propto 1/k_0 C$ \cite{Droulias2023b}. In Fig.\,\ref{fig:fig09}(b), the RIS of Fig.\,\ref{fig:fig09}(a) is now utilized to produce a Bessel-beam with $C=0.0125$, which propagates with main lobe of practically constant width, for an extended distance. Such beams are ideal for ensuring constant received power along straight paths, a key element for energy efficient communications and wireless power transfer \cite{Stratidakis2024}.\\
\indent Last, by breaking the radial symmetry of the beam upon substituting $\rho\rightarrow x$ in \eqref{Eq:Eq16}, we can generate beams that propagate on bent trajectories. For example, for
\begin{subequations}
    \begin{align}
        &a=-\frac{4}{3}k_0 \sqrt{\beta},\\
        &\gamma=3/2,
    \end{align}
    \label{Eq:EqADV4}
\end{subequations}
the reflected beam belongs to the general class of Airy beams, which evolve along the parabolic trajectory $x=\beta z^2$ \cite{Niu2017,Froehly2011} (typically, Airy beams additionally require a tailored amplitude profile). This possibility is demonstrated in Fig.\,\ref{fig:fig09}(c), where the RIS reconstructs the incident beam into a bending beam with $\beta=0.0025\,\mathrm{m}^{-1}$, which propagates along the analytically predicted trajectory marked with the dashed line. Such beams are ideal for blockage avoidance and for reaching users at locations beyond the LOS, i.e. users that not accessible with conventional beams that evolve along straight paths.
\section{Discussion and conclusion}
\noindent Future networks are envisioned to be equipped with multiple functionalities beyond conventional beamforming, to provide high quality of service in scenarios where the topology changes dynamically. In the examples demonstrated in this work, the RIS transfer function captures a specific configuration (link topology and associated beam properties) and is, hence, associated with a certain set of phase shifts. In future wireless communications the tunable features of the RIS are leveraged to serve dynamic scenarios with, possibly, very high mobility, and our framework also captures this tunability: a tunable RIS that is able to dynamically reconfigure the phase shifts serves $N$ different configurations and is characterized by a set of $T_{\mathrm{RIS},i}$, with $i = 1, 2,\dots, N$. Our framework is applied successively to the ensemble of all $N$ configurations. \\
\indent In this work, we investigated how a RIS can be operated as a spatial filter when it reflects incident waves. We showed that the impact of the RIS on the incident wave is to modify its $k$-content with direct consequences on crucial quantities, such as the power and spatial distribution of the reflected wave. We demonstrated how the RIS transfer function operates on the incident $k$-content. Based on our analytical approach, we showed how, by engineering the RIS shape and size, it is possible to manipulate the far-field distribution of the reflected wave to suppress unwanted interference, while concentrating the power in desired directions. We also demonstrated how our framework can be utilized to optimize the RIS phase shifts, and to tailor the incident beam's wavefront to generate beams that are capable of focusing, propagating with invariant profile and bending, beyond conventional beamforming. Our proposed framework provides the necessary insights into how the RIS response can be tailored to treat beams with nontrivial $k$-content for advanced operations, such as selective suppression of undesired incoming signals, sidelobe elimination of reflected waves and broadcasting to selected multiple users.
\appendix[ ]
\subsection{The RIS transfer function}
\label{AppendixA}
The operation of the RIS in real space is expressed as
\begin{equation}
    E_r(x,y) = \iint dx'dy' T_\mathrm{RIS}(x,y,x',y')E_i(x',y'),
    \label{Eq:EqA01}
\end{equation}
where the RIS transfer function is 
\begin{equation}
    T_\mathrm{RIS}(x,y,x',y')=\Gamma_0(x,y)e^{j(k_r-k_i)x} \delta(x-x',y-y').
    \label{Eq:EqA02}
\end{equation}
Direct substitution of \eqref{Eq:EqA02} into \eqref{Eq:EqA01} yields 
\begin{equation}
    E_r(x,y) = \Gamma_0(x,y)e^{j(k_r-k_i)x} E_i(x,y).
    \label{Eq:EqA03}
\end{equation}
To express $T_\mathrm{RIS}$ in the $k$-space, we apply the Fourier transform on $E_r$:
\begin{equation}
    \widetilde{E}_r(k_x,k_y) = \left(\frac{1}{2\pi}\right)^2\iint dxdy E_r(x,y) e^{-j(k_xx+k_yy)}.
    \label{Eq:EqA04}
\end{equation}
After using \eqref{Eq:EqA03} in \eqref{Eq:EqA04}, applying the Fourier transform of $E_i$ and rearranging the terms in the integral, we obtain
\begin{equation}
    \widetilde{E}_r(k_x,k_y) = \iint dk'_xdk'_y \widetilde{E}_i(k'_x,k'_y)\widetilde{\Gamma}_0(k_x-k_x'+k_i-k_r,k_y-k_y')
    \label{Eq:EqA05}
\end{equation}
and, hence
\begin{equation}
    \widetilde{T}_\mathrm{RIS}(k_x,k_y,k_x',k_y') = \widetilde{\Gamma}_0(k_x-k_x'+k_i-k_r,k_y-k_y').
    \label{Eq:EqA06}
\end{equation}
\subsection{The \textit{k}-content of typical beams}
\label{AppendixB}
\noindent The $k$-content of the incident beam depends on the particular antenna properties, ranging from beams with relatively narrow $k$-content that are reflected by the RIS similarly to plane waves, to beams with rich $k$-content that may interact with the RIS in a less straightforward manner. For typical beams generated by phased arrays and parabolic dishes, the $k$-content depends on the antenna size $D$, and is given by
\begin{equation}
    \frac{\Delta k}{k_0}\simeq 1.2\frac{\lambda}{D}.
    \label{Eq:EqB01}
\end{equation}
In view of \eqref{Eq:EqB01}, a relatively narrow $k$-content requires antenna apertures of at least $10\lambda$. For commonly used phased arrays, in particular, this corresponds to having at least 20 elements with $\lambda/2$ separation, bringing the RIS operation close to that of the plane wave limit. Next, we summarize the $k$-content of typical beams. 

For Gaussian beams, using  \eqref{Eq:EqEincGBk} we find that
\begin{equation}
    \frac{\Delta k}{k_0}\simeq \frac{\lambda}{2w_0}.
\end{equation}

For parabolic dish of diameter $D$ \cite{Kraus1988}, it follows that
\begin{equation}
    \frac{\Delta k}{k_0}\simeq \frac{1.22\lambda}{D}.
\end{equation}

For a phased array consisting of $N$ elements with inter-element distance $d$ \cite{Kraus1988}, we obtain
\begin{equation}
    \frac{\Delta k}{k_0}\simeq \frac{1.2\lambda}{Nd}.
\end{equation}
\subsection{Far-field distribution of received power}
\label{AppendixC}
Let us consider a point defined by the observation vector $\textbf{r} = x\hat{\textbf{x}}+y\hat{\textbf{y}}+z\hat{\textbf{z}}$ (or, in spherical coordinates, $\textbf{r} = r\sin{\theta}\cos{\varphi}\hat{\textbf{x}} + r\sin{\theta}\sin{\varphi}\hat{\textbf{y}} + r\cos{\theta}\hat{\textbf{z}}$). The power at the observation point that is captured by a receiver with antenna aperture $A_r$, is  given by $P_r = A_r \textbf{S}_r(\textbf{r}) \cdot \hat{\textbf{r}} $, or \cite{Droulias2023}
\begin{multline}
    P_r(\textbf{r}) = A_r \frac{k^2}{2Z_0} \Theta(\theta,\varphi) \left|\iint_{S} E_r(\textbf{r}') G(\textbf{r}-\textbf{r}')d\textbf{r}'\right|^2,
    \label{Eq:EqC01}
\end{multline}
where
\begin{align}
    \begin{split}
        \Theta(\theta,\varphi) = \sin^2{\varphi}(1+\cos{\theta}\cos{\theta_r})^2\\
        +\cos^2{\varphi}(\cos{\theta}+\cos{\theta_r})^2
    \end{split}
    \label{Eq:EqC02}
\end{align}
and
\begin{align}
        G(\textbf{r}-\textbf{r}') = \frac{e^{-j k |\textbf{r}-\textbf{r}'|}}{4\pi |\textbf{r}-\textbf{r}'|}
        \label{Eq:EqC03}
\end{align}
is the Green’s function for the Helmholtz equation. 
In the far-field, we may use the approximation
\begin{align}
        G(|\textbf{r}-\textbf{r}'|) \simeq \frac{e^{-j k r}}{4\pi r}e^{-j \textbf{k} \textbf{r}'},
            \label{Eq:EqC04}
\end{align}
and \eqref{Eq:EqC01} takes the form
\begin{multline}
    P_r(\textbf{r}) = A_r \frac{k^2}{2Z_0} \Theta(\theta,\varphi) \frac{1}{(4\pi r)^2} \left|\iint_{S} E_r(\textbf{r}') e^{-i \textbf{k} \textbf{r}'}d\textbf{r}'\right|^2
    \label{Eq:EqC05}
\end{multline}
or
\begin{equation}
    P_r(\textbf{r}) = A_r \frac{k^2}{2Z_0} \Theta(\theta,\varphi) \frac{\left|\widetilde{E}_r(k_x,k_y)\right|^2}{(4\pi r)^2},
    \label{Eq:EqC06}
\end{equation}
where the Fourier transform of $E_r$ is
\begin{equation}
    \widetilde{E}_r(k_x,k_y)\equiv \iint_{S}E_r(\textbf{r}') e^{-j \textbf{k} \textbf{r}'}d\textbf{r}'.
    \label{Eq:EqC07}
\end{equation}
\subsection{Propagation in free-space}
\label{AppendixD}
Any beam can be expressed as a linear superposition of plane waves as
\begin{equation}
    E_\mathrm{beam}(x,y) = \iint dk_xdk_y \widetilde{E}_\mathrm{beam}(k_x,k_y) e^{j(k_xx+k_yy)},
    \label{Eq:EqD01}
\end{equation}
where $\widetilde{E}_\mathrm{beam}(k_x,k_y)$ denotes the $k$-dependent weights of the individual plane waves. Note that this is simply the Fourier transform of $E_\mathrm{beam}(x,y)$. Each plane wave propagates acquiring a phase $e^{jk_zz}$, where $k_z=\sqrt{k^2_0-k^2_x-k^2_y}$ is the transverse wavenumber that limits the propagating $k$-components within the range $k_x^2+k_y^2/k_0^2<1$. Because beam propagation is equivalent to the propagation of the individual plane waves, the beam propagates in the $k$-space as $E(k_x,k_y)e^{jk_zz}$, i.e., the $k$-content of the beam acquires a global phase without undergoing other changes in magnitude and phase (except for the case of lossy atmosphere or turbulent conditions, which are cases beyond the scope of this work). \\
\indent Taking into account that \eqref{Eq:EqD01} is the inverse Fourier transform of the beam, beam propagation can be written concisely as
\begin{equation}
    E_\mathrm{beam}(x,y,z+\delta z) = \mathcal{FT}\,^{-1}\{\,\mathcal{FT}\,[E_\mathrm{beam}(x,y,z)]e^{jk_z\delta z}\},
    \label{Eq:EqD02}
\end{equation}
where \(\mathcal{FT}\) denotes the Fourier transform and \(\mathcal{FT}\)\,$^{-1}$ its inverse. Due to linearity, if the $k$-content is known at a certain propagation step, the beam can be reconstructed anywhere in real space.

\subsection{RIS transfer function for spatial filtering}
\label{AppendixE}
\indent To eliminate all waves arriving with $k\neq k_i$, let us use a filter function of the form $\exp\left(-(k_x-k_i)^2/2 k_F^2\right)$, where $2\sqrt{2\ln{2}}k_F$ is the Full Width at Half Maximum (FWHM) of the filter. This filter is applied to the incident waves, which are subsequently shifted by $k_r-k_i$ and, hence, the RIS transfer function reads
\begin{equation}
        \widetilde{T}_\mathrm{RIS}=\Gamma_0 e^{-\frac{(k_x'-k_i)^2}{2 k_F^2}} \delta(k_x-k_x'+k_i-k_r,k_y-k_y').
    \label{Eq:EqE01}
\end{equation}
Inserting \eqref{Eq:EqE01} into \eqref{Eq:EqErINT} results in
\begin{equation}
    \widetilde{E}_r(k_x,k_y) = \Gamma_0 e^{-\frac{(k_i-k_n)^2}{2k_F^2}}\widetilde{E}_i(k_x+k_i-k_r,k_y).
    \label{Eq:EqE02}
\end{equation}

To express $\widetilde{T}_\mathrm{RIS}$ in real space, we apply the inverse Fourier transform on $\widetilde{E}_r$ given by \eqref{Eq:EqE02}:
\begin{equation}
    E_r(x,y) = \iint dk_xdk_y \widetilde{E}_r(k_x,k_y) e^{j(k_xx+k_yy)},
    \label{Eq:EqE03}
\end{equation}
which leads to the final result
\begin{equation}
        T_\mathrm{RIS}=\Gamma_0 \frac{k_F}{\sqrt{2\pi}} \exp \left(-\frac{k_F^2}{2}(x-x')^2 + j(k_r x - k_i x')\right).
    \label{Eq:EqE04}
\end{equation}

\subsection{Analytical model for RIS element reflectivity}
\label{AppendixF}
To express the RIS element reflection coefficient analytically, we first replace the RIS with a thin radiating sheet \cite{Droulias2023} that is characterized by the electric and magnetic conductivities, $\sigma_e$ and $\sigma_m$, respectively. For a reflective sheet with zero transmission it is required that $2/\sigma_e(f)=\sigma_m(f)/2\equiv Z(f)$, where $Z(f)$ is the sheet impedance and $f$ is the operation frequency. The sheet reflectivity is given by
\begin{equation}
    r(f)=\frac{Z(f)-Z_0}{Z(f)+Z_0},
    \label{Eq:RISrZ}
\end{equation}
where $Z_0$ is the free-space wave impedance. Using the normalized conductivity $\hat{\sigma}_e=Z_0\sigma_e/2$, $r(f)$ is written concisely as
\begin{equation}
    r(f)=\frac{1-\hat{\sigma}_e(f)}{1+\hat{\sigma}_e(f)}.
    \label{Eq:RISrS}
\end{equation}
Next, we use a Lorentzian function to model the resonant conductivity $\hat{\sigma_e}(f)$, as
\begin{equation}
    \hat{\sigma}_e(f)=\frac{ja_ef}{f^2_0-f^2+j\gamma_ef},
    \label{Eq:sLorentzian}
\end{equation}
where $f_0$ is the resonance frequency, $\gamma_e$ the damping rate, and the parameter $a_e$ controls the resonance strength. We can now attribute the local sheet reflectivity to the response of each RIS element and, hence, we may write $\Gamma_\mathrm{UC}(f)\approx r(f)$. Substituting \eqref{Eq:sLorentzian} into \eqref{Eq:RISrS} leads to
\begin{equation}
    \Gamma_\mathrm{UC}(f)=1-\frac{2a_ef}{(a_e+\gamma_e)f + j(f^2-f_0^2)},
    \label{Eq:RfitLorentzian}
\end{equation}
the amplitude and phase of which, are given explicitly in \eqref{Eq:GammaUCabs} and \eqref{Eq:GammaUCarg}, respectively.

\subsection{RIS phase optimization scheme}
\label{AppendixG}
To solve the optimization problem (25) we write as $\Gamma = (\Gamma_{0,\mathrm{theory}}+\delta\Gamma_0)e^{j(\phi_{\mathrm{theory}}+\delta\phi)}$ the reflection coefficient to be determined, where $\Gamma_{0,\mathrm{theory}}$, $\phi_\mathrm{theory}$ are the theoretical reflection amplitude and phase, respectively, and $\delta\Gamma_0$, $\delta\phi$ are the detunings from the theoretical values. The corresponding RIS transfer function is expressed as
\begin{equation}
    T_\mathrm{RIS}(x,y,x',y')=\Gamma(x,y) \delta(x-x',y-y').
    \label{Eq:EqG01}
\end{equation}
Using \eqref{Eq:EqA01}, the reflected wave is written as
\begin{equation}
    E_r(x,y) = \iint dx'dy' T_\mathrm{RIS}(x,y,x',y')E_i(x',y'),
    \label{Eq:EqG02}
\end{equation}
and, using the Fourier transform of $E_{r}(x,y)$, the corresponding footprint $\widetilde{E}_{r}(k_x,k_y;\Gamma)$ is expressed as
\begin{equation}
    \widetilde{E}_r(k_x,k_y;\Gamma) = \left(\frac{1}{2\pi}\right)^2\iint dxdy \Gamma(x,y) E_i(x,y) e^{-j(k_xx+k_yy)}.
    \label{Eq:EqG03}
\end{equation}
For completeness we repeat here the theoretical footprint
\begin{equation}
    \widetilde{E}_r(k_x,k_y;\Gamma_\mathrm{theory}) = \left(\frac{1}{2\pi}\right)^2\iint dxdy \Gamma_\mathrm{theory}(x,y) E_i(x,y) e^{-j(k_xx+k_yy)},
    \label{Eq:EqG04}
\end{equation}
which is associated with the RIS reflection coefficient \eqref{Eq:EqREFLCOEFF}, with $\Gamma_0(x,y)=1\equiv \Gamma_{0,\mathrm{theory}}$ and $\phi(x,y) = k_0(\sin{\theta_r}-\sin{\theta_i})x \equiv \phi_\mathrm{theory}$. \\
\indent To numerically perform the integration in \eqref{Eq:EqG03} and \eqref{Eq:EqG04}, we first discretize the RIS surface into a grid of $N_x\times N_y$ points with periodicity $l_x$ and $l_y$ along the $x$ and $y$ axis, respectively. Next we discretize \eqref{Eq:EqG03}, \eqref{Eq:EqG04} on the RIS grid and perform the optimization scheme \eqref{Eq:EqOPT} using the $fmincon$ function in Matlab, which is designed to find the minimum of constrained nonlinear multivariable functions. We initialize the $fmincon$ function with the theoretical reflection coefficient, and execute the scheme to calculate the unknown $\delta\Gamma_0$, $\delta\phi$. 

\bibliographystyle{IEEEtran}
\bibliography{IEEEabrv,main}
\end{document}